\documentclass[12pt]{article}
\usepackage{graphicx}
\usepackage{hep}
\usepackage{amsmath}
\usepackage{amssymb}
\usepackage{cite}
\newcommand{\sul}{\ensuremath{\tilde{u}_\textrm{L}}}
\newcommand{\sur}{\ensuremath{\tilde{u}_\textrm{R}}}
\newcommand{\sdl}{\ensuremath{\tilde{d}_\textrm{L}}}
\newcommand{\sdr}{\ensuremath{\tilde{d}_\textrm{R}}}

\newcommand{\msdl}{\ensuremath{m_{\sdl}}}
\newcommand{\msdr}{\ensuremath{m_{\sdr}}}
\newcommand{\msul}{\ensuremath{m_{\sul}}}
\newcommand{\msur}{\ensuremath{m_{\sur}}}

\newcommand{\msusy}{\ensuremath{M_{\mathrm{SUSY}}}}
\newcommand{\deltatt}{\ensuremath{\delta_{23}}}
\newcommand{\hqq}{\ensuremath{h\to q\,{q'}}}

\newcommand{\hbs}{\ensuremath{h\to b\,s}}
\newcommand{\hzbs}{\ensuremath{h^0\to b\,s}}

\newcommand{\htc}{\ensuremath{h\to t\,c}}

\newcommand{\Hztc}{\ensuremath{H^0\to t\,c}}

\newcommand{\sigmapphqq}{\ensuremath{\sigma(pp\to \hqq)}}

\newcommand{\sigmapphbs}{\ensuremath{\sigma(pp\to \hbs)}}
\newcommand{\sigmapphzbs}{\ensuremath{\sigma(pp\to \hzbs)}}

\newcommand{\sigmapphtc}{\ensuremath{\sigma(pp\to \htc)}}

\newcommand{\sigmappHztc}{\ensuremath{\sigma(pp\to \Hztc)}}

\newcommand{\Ghbs}{\ensuremath{\Gamma(h\to q\,{q'})}}

\newcommand{\bsg}{\ensuremath{b\to s\gamma}}
\newcommand{\Bbsg}{\ensuremath{B(b\to s\gamma)}}
\newcommand{\GeV}{\mbox{\,GeV}}
\newcommand{\TeV}{\mbox{\,TeV}}

\newcommand{\sbottom}{\ensuremath{\tilde{b}}}
\newcommand{\sstrange}{\ensuremath{\tilde{s}}}
\newcommand{\squark}{\ensuremath{\tilde{q}}}

\newcommand{\pb}{\mbox{\,pb}}
\newcommand{\fb}{\mbox{\,fb}}
\def\Title#1{\begin{center} {\Large\bf #1 } \end{center}}
\def\Author#1{\begin{center}{ \sc #1} \end{center}}
\def\Address#1{\begin{center}{ \it #1} \end{center}}

\newcommand\pubblock{\rightline{\begin{tabular}{l} \pubnumber\\
        \pubdate\\ \hepnumber \end{tabular}}}
\newcommand\pubnumber{UAB-FT-583\\UB-ECM-PF-05/15}
\newcommand\pubdate{July 2005}
\newcommand\hepnumber{hep-ph/0508043}

\newcommand{\GuaschNPo}{Guasch:1999jp}
\newcommand{\BejarNPo}{Bejar:2000ub}
\newcommand{\BejarNPt}{Bejar:2003em}
\newcommand{\BejarJHEP}{Bejar:2004rz}
\newcommand{\BejarRADCOR}{Bejar:2001sj}
\newcommand{\CGGJS}{Coarasa:1996qa}

\newcommand{\PDG}{Eidelman:2004wy}
\newcommand{\SUSY}{Nilles:1984ex,Haber:1985rc,Lahanas:1987uc,Ferrara87}

\newcommand{\LHC}{Atlas,CMS}
\newcommand{\GuaschSola}{Guasch:1999jp,Bejar:2001sj}
\newcommand{\Dabels}{Yamada:1994kj,Chankowski:1994er,Dabelstein:1995hb,Dabelstein:1995js}

\newcommand{\bsgexp}{Alam:1995aw,Barate:1998vz,Ahmed:1999fh,Abe:2001hk,Chen:2001fj,Aubert:2002pd}

\begin{document}

\pubblock

\vfill
\def\thefootnote{\fnsymbol{footnote}}
\Title{Production and FCNC decay of supersymmetric Higgs bosons
into heavy quarks in the LHC}
\Author{Santi
  B{\'e}jar$^{a,b}$, Jaume Guasch$^{c}$, Joan Sol{\`a}$^{c,b}$}
\Address{$^{a}$\textsl{Grup de F{\'\i}sica Te{\`o}rica,
    Universitat Aut{\`o}noma de Barcelona,\\
    E-08193, Bellaterra,
    Barcelona, Catalonia, Spain\\ }
  $^{b}$\textsl{Institut de F{\'\i}sica d'Altes Energies, Universitat
    Aut{\`o}noma de Barcelona,\\
    E-08193, Bellaterra, Barcelona, Catalonia, Spain\\ }
  $^{c}$\textsl{Departament d'Estructura i Constituents de la
    Mat{\`e}ria, Universitat de Barcelona,\\
    E-08028, Diagonal 647, Barcelona, Catalonia, Spain}
} \vspace{1cm}




\begin{abstract}
We analyze the production and subsequent decay of the neutral
MSSM Higgs bosons ($h\equiv h^0,\ H^0,\ A^0$) mediated by flavor
changing neutral currents (FCNC) in the LHC collider. We have
computed the $h$-production cross-section times the FCNC branching
ratio, $\sigma(pp\to h\to q{q}')\equiv\sigma(pp\to h) \times
B(h\to q{q}')$, in the LHC focusing on the strongly-interacting
FCNC sector. Here $qq'$ is an electrically neutral pair of quarks
of different flavors, the dominant modes being those containing a
heavy quark:  $tc$ or $bs$. We determine the maximum production
rates for each of these modes and identify the relevant regions
of the MSSM parameter space, after taking into account the severe
restrictions imposed by low energy FCNC processes. The analysis
of $\sigma(pp\to h\to q{q}')$ singles out regions of the MSSM
parameter space different from those obtained by maximizing only
the branching ratio, due to non-trivial correlations between the
parameters that maximize/minimize each isolated factor. The
production rates for the $bs$ channel can be huge for a FCNC
process ($0.1-1 \pb$), but its detection can be problematic. The
production rates for the $tc$ channel are more modest
($10^{-3}-10^{-2}\pb$), but its detection should be easier due to
the clear-cut top quark signature. A few thousand $tc$ events
could be collected in the highest luminosity phase of the LHC,
with no counterpart in the SM.
\end{abstract}

\baselineskip=5.6mm
\def\thefootnote{\arabic{footnote}}
\newpage

\section{Introduction}
The search for physics beyond the Standard Model (SM) is a very
relevant, if not the most important, endeavor within the big
experimental program scheduled for the forthcoming Large Hadron
Collider (LHC) experiment at CERN~\cite{\LHC}. There are several
favorite searching lines on which to concentrate, but undoubtedly
the most relevant one (due to its central role in most extensions
of the SM) is the physics of the Higgs boson(s) with all its
potential physical manifestations. Supersymmetry
(SUSY)~\cite{\SUSY} is certainly related to Higgs boson physics,
and at the same time it may convey plenty of additional
phenomenology. Ever since its inception, SUSY has been one of the
most cherished candidates for physics beyond the SM, and as such
it will be scrutinized in great detail at the LHC. It is no
exaggeration to affirm that the LHC will either prove or disprove
the existence of SUSY, at least in its most beloved low-energy
realization, namely the one which is needed to solve the
longstanding naturalness problem in the Higgs sector of the SM
\,\cite{Haber:1985rc}. Indeed, if SUSY is realized around the TeV
scale, the LHC experiments shall be able to directly produce the
SUSY particles for masses smaller than a few
TeV\,\cite{Weiglein:2004hn,Degrassi:2004ed}. On the other hand,
the presence of SUSY may also be tested indirectly through the
quantum effects of the supersymmetric particles. For one thing it
has been known since long ago that SUSY particles may produce
large virtual effects on Higgs boson observables\,\footnote{See
e.g.\,\cite{Guasch:1995rn,Coarasa:1996yg,Coarasa:1996qa,Guasch:1997jc,Coarasa:1997ky,
Coarasa:1999da,Coarasa:1999db,Coarasa:1999wy,Belyaev:2001qm,Belyaev:2002eq,
Belyaev:2002sa,Guasch:2001wv,Guasch:2003cv} and references
therein. For a review see e.g. \cite{Carena:2002es}.}.

Within the general strategy based on detecting indirect effects
of the new physics, Flavor Changing Neutral Currents (FCNC) play a
very special role. In the SM they are completely absent at the
tree-level~\cite{Glashow:1970gm}. At one loop, however, the FCNC
effects are possible in the SM, but then the contributions from
the new particles enter on equal footing with those from SM
particles. Therefore, it is not inconceivable that in certain
regions of the parameter space the new physical effects may well
dominate the SM contributions, thus providing a unique signature
of physics beyond the SM. This is particularly so when the SM
one-loop effects, even though non-vanishing, turn out to be highly
suppressed. In such situations the sole observation of these FCNC
processes would be instant evidence of new physics. A most
dramatic example of this kind of appealing scenarios occurs within
the FCNC physics of the SM Higgs boson ($H_{\rm SM}$)
interactions with quarks. For example, the FCNC vertex $H_{\rm
SM}tc$ may lead (at one loop) to such rare decays as
$t\rightarrow H_{\rm SM}\,c$ or $H_{\rm SM}\rightarrow t\,c$,
depending on the mass of $H_{\rm SM}$ (At the moment both
possibilities are still open because the LEP bounds amount to a
mass for $H_{\rm SM}$ above $114.4\GeV$\,\cite{\PDG}.). Both of
these modes are extremely suppressed at one loop, with branching
ratios of order $10^{-14}$ or less, hence $10$ orders of magnitude
below other more conventional (and relatively well measured) FCNC
processes like $b\rightarrow s\gamma$\,\cite{\PDG}. The rareness
of the FCNC Higgs boson decay modes is borne out by direct
calculation\,\cite{Mele:1998ag}, and it can also be understood
from physical arguments based on dimensional analysis, power
counting, CKM matrix elements and dynamical
features\,\cite{\BejarJHEP,Bejar:2003em,\BejarNPo}. Therefore,
these processes are an ideal laboratory to look for non-standard
interactions superimposed onto the SM ones.  Similar
considerations apply to the FCNC processes associated to the $Hbs$
vertex, but in this case it is more difficult to pin down the
phenomenological signatures. Some work along these lines has
already been done, both in the MSSM~\cite{Guasch:1999jp,Guasch:1997kc,Guasch:1999ve,
\BejarRADCOR,\BejarJHEP,Curiel:2002pf,Demir:2003bv,Curiel:2003uk,Heinemeyer:2004by}
and in the general two-Higgs-doublet model
(2HDM)\,\cite{Bejar:2003em,\BejarNPo,Bejar:2001sj,Arhrib:2004xu},
and also in other extensions of the SM-- see
\,\cite{Aguilar-Saavedra:2004wm} for a review. Up to now, the
main effort has been concentrated in computing the FCNC decay
modes at one-loop within the new physics, and also in getting a
realistic estimate of the maximum branching ratios expected. It
is not enough to compute the FCNC branching ratios in, say the
MSSM, and then evaluate them in some favorable region of the
parameter space, for one has to preserve at the same time the
stringent bounds on other observables in which the same physics
can be applied, like the aforementioned low-energy $b\rightarrow
s\gamma$ decay. This kind of correlated study was done very
carefully in Ref.~\cite{\BejarJHEP} for the specific Higgs boson
FCNC decays into bottom quarks within the MSSM, $h\rightarrow
b\,s\ (h=h^0,H^0,A^0)$. In this paper we extend the latter work by
computing also the top quark Higgs boson FCNC decay modes of the
heavy MSSM Higgs bosons, $h\rightarrow t\,c\ (h=H^0,A^0)$ under
the same restrictions (recall that $h^0$ cannot participate in
this decay because $\mh<m_t$ in the
MSSM\,\cite{Carena:2002es}). Furthermore, in this work we carry
out an additional step absolutely necessary to make contact with
experiment, namely we combine the FCNC decay branching ratios of
the MSSM Higgs bosons (into both top and bottom quarks) with
their MSSM production cross-sections in order to estimate the
maximum number of FCNC events expected at LHC energies and
luminosities. Only in this way one can assess in a practical way
the probability of detecting such processes at the LHC. While
this computation is already in the literature for the general
2HDM\,\cite{\BejarNPt}, to the best of our knowledge the
corresponding calculation in the MSSM case is not available. In
this paper we perform this calculation and compare the FCNC
results obtained for the MSSM and 2HDM scenarios.

The relevant observable quantity on which we shall focus hereafter
is the cross-section for the production of electrically neutral
pairs of heavy quarks of different flavors at the LHC, whose
origin stems from the FCNC decays of the neutral Higgs bosons of
the MSSM, $h=h^0,\,H^0,\,A^0$\,\cite{Hunter,Carena:2002es}. Thus,
we aim at the quantity
\begin{eqnarray}
        \sigmapphqq
        &\equiv&
        \sigma(pp\to h X)B(\hqq)\equiv \sigma(pp\to h X) \frac{\Gamma(\hqq)}{\Gamma(h\to X)}\nonumber\\
        &\equiv& \sigma(pp\to h X) \frac{\Gamma(h\to
          q\,\bar{q'}+\bar{q}\,q')}{\sum_i \Gamma(h\to X_i)}\ \ \
        (qq'\equiv bs \mbox{ or } tc )\,.
    \label{eq:hqq-def}
\end{eqnarray}
Here $\Gamma(h\to X)$ is the -- consistently computed -- total
width in each case. In order to asses the possibility to measure
these processes at the LHC, we have performed a scan of the MSSM
parameter space to find the maximum possible value of the
production rates~(\ref{eq:hqq-def}) under study. The computation
of the combined production rate is necessary, since the
correlations among the different factors are important. For
example, in Ref.~\cite{\BejarJHEP} it was shown that the maximum
branching ratio for the lightest MSSM Higgs boson, $B(h^0\to
b\bar{s})$, is obtained in the regions of the parameter space
where the coupling $h^0\,b\,\bar{b}$ is strongly suppressed by
quantum effects. On the other hand, the associated production
$\sigma(pp\to h^0 b\bar{b})$ is one of the leading processes for
the production of the lightest MSSM Higgs boson. It is clear then
that, in the regions where $B(h^0\to b\bar{s})$ is largest,
$\sigma(pp\to h^0 b\bar{b})$ will be suppressed. Therefore, the
maximum FCNC production rate at the LHC can only be obtained by
the combined analysis of the two relevant factors in
(\ref{eq:hqq-def}) (viz. the branching ratio and the Higgs boson
production cross-section). We will see that the effects from each
factor are different in different regions of the parameter space.
Moreover, the realistic production FCNC rates (\ref{eq:hqq-def})
in the MSSM parameter space can be obtained only by including the
restrictions imposed by the simultaneous analysis of the
branching ratio of the low-energy process $b\to s\,\gamma$, whose
range of values is severely limited by experiment~\cite{\bsgexp}.
As in Ref.\,\cite{\BejarJHEP}, in this paper we limit ourselves to
supersymmetric FCNC interactions mediated by the strongly
interacting sector of the MSSM, i.e. the SUSY-QCD
flavor-violating interactions induced by the gluinos. The
corresponding analysis for the electroweak supersymmetric FCNC
effects requires a lengthy separate presentation, and will be
reported elsewhere\,\cite{FCNCEW}.

The paper is organized as follows. In Section 2 we describe the
general setting for our numerical analysis. In Section 3 we
present the LHC production rates of Higgs boson decaying into
bottom quarks through supersymmetric FCNC interactions. In Section
4 we present the corresponding FCNC rates for the top quark
channel. Finally, in Section 5 we compare the MSSM results with
the 2HDM results, and deliver our conclusions.

\section{General setting for the numerical analysis}
\label{sec:numerical-analysis}

To compute the full one-loop value of the FCNC cross-sections
$\sigmapphqq$ for the three MSSM Higgs bosons ($h=
h^0,\,H^0,\,A^0$) we shall closely follow the notation and methods
of Refs.\,\cite{\BejarNPo,\BejarNPt,\BejarJHEP,\GuaschNPo}. We
refer the reader to these references for the technical details. In
particular, a thorough exposition of the relevant interaction
Lagrangians and similar set of Feynman diagrams for the FCNC
interactions is provided in \cite{\GuaschNPo}. See also
\cite{Hunter,Carena:2002es} for basic definitions in the MSSM
framework and \cite{\CGGJS} for detailed computational techniques
and further illustration of the supersymmetric enhancement
effects in other relevant Higgs boson processes.  There is no need
to go through these lengthy details here and in what follows we
shall limit ourselves to present the final results of our
numerical analysis together with a detailed discussion,
interpretation and phenomenological application.

We have performed the calculations with the help of the numeric
programs \texttt{HIGLU},
\texttt{PPHTT}~\cite{Spira,Spira:1995mt,Spira:1995rr,Spira:1997dg}
and LoopTools~\cite{Hahn:1998yk,LTuser,vanOldenborgh:1989wn}. The
calculation must obviously be finite without renormalization, and
indeed the cancellation of UV divergences using either
dimensional regularization or dimensional reduction -- the two
methods giving the same results here -- in the total amplitudes
was verified explicitly. In the following we will detail the
approximations used in our computation:
\begin{itemize}
\item We include the full one-loop SUSY-QCD contributions to the FCNC partial
  decay widths $\Ghbs$ in the observable (\ref{eq:hqq-def}).
\item We assume that FCNC mixing terms appear only in the
  {LH-chiral} sector of the $6\times 6$ squark mixing matrix. Therefore, this matrix has only
  non-diagonal blocks in the LH-LH sector. This is
  the most natural assumption from the theoretical point of view\,\cite{Duncan:1983iq}, and,
  moreover, it was proven in Ref.~\cite{\GuaschSola} that the presence of FCNC terms in the
  {RH-chiral sector} would enhance the partial widths by a factor two
  at most -- not an order of magnitude.
\item The Higgs sector parameters (masses and CP-even mixing angle
  $\alpha$) {have} been treated using the leading $\mt$ and $\mb\tb$
  approximation  to the one-loop result~\cite{\Dabels}. For comparison, we also
  perform the analysis using the tree-level approximation.
\item The Higgs bosons total decay widths $\Gamma(h\to X)$ are computed
  at leading order, including all the relevant channels: $\Gamma(h\to
  f\bar{f},ZZ,W^+W^-,gg)$. The off-shell decays
  $\Gamma(h\to ZZ^*,W^{\pm}W^{\mp*})$ have also been
  included. This is necessary to consistently compute the
  total decay width of $\Gamma(h^0\to X)$ in regions of the parameter
  space where the maximization of the cross-section (\ref{eq:hqq-def})
  is obtained at the expense of greatly diminishing the partial decay widths of the two-body
  process $h^0\rightarrow b\bar{b}$ (due to dramatic quantum effects that may reduce
  the CP-even mixing angle  $\alpha$ to small values~\cite{Carena:2002qg}).
  The one-loop decay rate $\Gamma(h\to gg)$
  has been taken from~\cite{Spira:1995rr} and the off-shell decay
  partial widths have been recomputed explicitly and found perfect agreement with the old
  literature on the subject\,\cite{Keung:1984hn}.
\item The MSSM Higgs boson production cross-sections at the LHC have been computed
using the programs  \texttt{HIGLU 2.101} and \texttt{PPHTT
1.1}~\cite{Spira,Spira:1995mt,Spira:1995rr,Spira:1997dg}. These
programs include the following channels: gluon-gluon fusion
$\sigma(pp(gg)\to h)$, associated production with top-quarks
$\sigma(pp \to h t\bar{t})$ and associated production with
bottom-quarks $\sigma(pp\to h b\bar{b})$. In order to have a
consistent description, we have used the leading order
approximation for all channels. The QCD renormalization scale is
set to the default values for each program, namely $\mu_0=\Mh$ for
\texttt{HIGLU} and  $\mu_0=(\Mh+2 M_Q)/2$ for \texttt{PPHTT}. We
have used the set of CTEQ4L Parton Distribution
Functions~\cite{Lai:1996mg}.
\end{itemize}
Running quark masses ($m_q(Q)$) and strong coupling constants
($\alpha_s(Q)$) are used throughout, with the renormalization scale set
to the decaying Higgs boson mass in the decay processes.
More details are given
below, as necessary.

Using this setup, we have performed a maximization of the FCNC
cross-section, Eq.~(\ref{eq:hqq-def}), in the MSSM parameter space
with the following restrictions on the parameters:
\begin{equation}
    \begin{array}{c|c|c}
        qq'&bs&tc\\\hline\\
        \deltatt&\multicolumn{2}{c}{<10^{-0.09} \simeq 0.81}\\
        \tan\beta & 50 & 5\\
        A_t&-300 \GeV&|A_t|\leq3\msusy\\
        A_b&|A_b|\leq3\msusy&300 \GeV\\
        \mu&\multicolumn{2}{c}{(-1000 \cdots 1000) \GeV}\\
        m_{\tilde{q_i}}&\multicolumn{2}{c}{\msdl=\msdr=\msur=\mg\equiv\msusy}\\
        \msusy&\multicolumn{2}{c}{(150 \cdots 1000) \GeV}\\
        \mA& \multicolumn{2}{c}{(100\cdots 1000)\GeV}\\
         M_{\tilde{q_i}}& \multicolumn{2}{c}{2\,M_{\tilde{q_i}}>\mH+
         50\GeV}\\
         &
         \multicolumn{2}{c}{M_{\tilde{q}_i}+M_{\tilde{q}_j}>\mA+
         50\GeV \, \, (i\neq j)}\\
        \\\hline
    \end{array}
    \label{eq:scan-parameters}
\end{equation}
Here $m_{\tilde{q_i}}$ are the  LH-{chiral} and RH-chiral squark
soft-SUSY-breaking mass parameters, $m_{\tilde{q}_{L,R}}$, common
for the three generations; $M_{\tilde{q_i}}$ are the physical
masses of the squarks, and $\Mh$ is the mass of the decaying
Higgs boson $h=h^0,H^0,A^0$. These masses are fixed at the
tree-level by the values of $(\tb,\mA)$ and the SM gauge boson
masses and couplings\,\cite{Hunter}. Due to the structure of the
Yukawa couplings in the MSSM, the value of $\tb$ is fixed at a
high (small) value for the bottom (top) quark channel as
indicated. The parameter $\mA$ (the mass of the CP-odd Higgs
boson) is assumed to vary in the range indicated in
(\ref{eq:scan-parameters}). At one loop these masses receive
corrections from the various SUSY fields, and therefore depend on
the values of the remaining parameters in
Eq.\,(\ref{eq:scan-parameters}). The characteristic SUSY mass
scale $\msusy$ defines the typical mass of the squark and gluino
masses\,\footnote{Our programs are able to deal with completely
arbitrary masses for each squark, but we are forced to make some
simplifications in order to provide a reasonable analysis within
a manageable total CPU time, see below.}. The rest of the
parameters of the squark sector are determined by this setup. For
instance, by $SU(2)$ gauge invariance we have $\msul=\msdl$.
Following the same notation as in~\cite{\GuaschNPo}, the
  parameter $\delta_{23}$  represents the mixing between the second
  and third generation of LH-chiral squarks. Let us recall its definition:
\begin{equation}\label{delta23}
    \delta_{23}\equiv \frac{m^2_{\sbottom_L
        \sstrange_L}}{m_{\sbottom_L} m_{\sstrange_L}}\,,
\end{equation}
$m^2_{\sbottom_L \sstrange_L}$ {being} the non-diagonal term in
the squark mass matrix squared mixing the second and third
generation of LH-chiral squarks -- and an equivalent definition
for the up-type quarks.\footnote{Recall that the $\delta_{ij}$
  parameters in the up-sector are related to the corresponding
  parameters in the down-sector by the Cabibbo-Kobayashi-Maskawa matrix,
  see e.g.\cite{Gabbiani:1996hi,Misiak:1997ei}.}
The parameter $\delta_{23}$ is a fundamental quantity in our
analysis as it determines the strength of the tree-level FCNC
interactions induced by the supersymmetric strong interactions,
which are then transferred to the loop diagrams of the Higgs
boson FCNC decays in Eq.\,(\ref{eq:hqq-def}). The last two
restrictions in Eq.~(\ref{eq:scan-parameters}) ensure that the
(heavy) Higgs boson decay channels into a pair of squarks are
kinematically forbidden. We have checked explicitly for some of
the heavy Higgs boson channels ($h^0$ can never do it in practice)
that we obtain the same results if we remove these conditions and
include  the partial widths $\Gamma(h\to \tilde{q}\tilde{q}^*)$
in the denominator of (\ref{eq:hqq-def}). Strictly speaking this
condition could be implemented, in the case of the $H^0$ boson,
by just requiring $\mH<2\,M_{\tilde{q_i}}$, but we have made it
stronger by including an additive term. This term is arbitrary
(provided it is not very small) and acts as a buffer, namely it
impedes that by an appropriate choice of the squark masses we can
approach arbitrarily close the threshold from above, and
therefore avoids artificial enhancement effects in our loop
calculations (see below). Similarly, for the CP-odd Higgs boson,
the condition expressed in (\ref{eq:scan-parameters}) ensures
that we avoid a similar kind of enhancement. In this case,
however, the condition is a bit different because the
$A^0\,\tilde{q} \tilde{q}^*$ vertex can only exist with squarks
of different chirality types ($A^0\,\tilde{q_L} \tilde{q_R}^*$)
or, equivalently, with different mass eigenstates
($A^0\,\tilde{q_i} \tilde{q_j}^*$).
We have used fixed values for the soft-SUSY-breaking trilinear
couplings $A_t$ and $A_b$ for the $bs$ and $tc$ channels
respectively. Our results are essentially independent of these
values and their signs.

The task of scanning the MSSM parameter space in order to
maximize $\sigmapphqq$ for the various Higgs bosons is quite
demanding and highly CPU-time consuming, even under the
conditions imposed in Eq.~(\ref{eq:scan-parameters}). As stated
in the introduction, our code includes also the restrictions on
the MSSM parameter space due to the experimental constraint on
$\Bbsg$, and therefore contains the full one-loop SUSY-QCD
amplitude for $b\rightarrow s\gamma$ constructed from the FCNC
interactions induced by the gluinos. The scan was carried out
with the help of two entirely different methods. In the first
method we used a systematic procedure based on dividing the
parameter subspace~(\ref{eq:scan-parameters}) into a lattice
which we filled with points distributed in a completely
homogeneous way. The second is a Monte-Carlo based method, first
proposed in~\cite{Brein:2004kh}. We have adapted the well-known
Vegas integration program\cite{Lepage:1977sw} to generate a
sufficient number of ``interesting'' points in our parameter
subspace. The total number of points used in this case was far
smaller than in the first method. Obviously the lattice procedure
gives more accurate results by increasing arbitrarily the total
number of points, but the CPU time becomes prohibitively long for
the whole analysis. This is so even after factoring out in a
suitable way the phase-space integrals of the Higgs boson
production processes, so that these integrals are computed only
once for every fixed Higgs boson mass and for all the MSSM points
of our scan in the parameter subspace (\ref{eq:scan-parameters}).
The second method is comparatively much faster, but it still
involves a quite respectable amount of CPU time for the whole
analysis. We found that the partial results obtained by the two
methods are compatible at the level of $10-20\%$. For the study
of our FCNC processes we consider that this level of accuracy
should be acceptable, and for this reason all of the plots that
we present in this work have been finally computed with the
Vegas-based procedure. This also explains the wiggling appearance
observed in the profiles of the curves presented in Sections 3-4.
For any given abscissa point in each one of these curves, the
corresponding value on the vertical (ordinate) axis is somewhere
within a band whose width lies around $10-20\%$ of the central
value.

A few words on the effects of the $\Bbsg$ constraint in our
analysis are now in order. The SUSY-QCD contribution to $\Bbsg$
can be quite large, in fact as large as the SM one, and with any
sign. This raises the possibility of ``fine-tuning'' between the
two type of contributions in certain (narrow) regions of the
parameter space. As a consequence we could highly optimize our
FCNC rates in these regions without being in conflict with the
experimentally measured $\Bbsg$ band. We have checked that in
these regions the number of  FCNC events can be artificially
augmented by one or two orders of magnitude. Our scanning
procedure indeed finds automatically these fine-tuning domains.
However, we have systematically avoided them in the presentation
of our analysis (for more details cf.~\cite{\BejarJHEP}). In all
of our plots, therefore, we show the results obtained for the
non-fine-tuned case only. We adopt $\Bbsg=(2.1-4.5)\times
10^{-4}$ as the experimentally allowed range within three standard
deviations~\cite{\PDG}.

\section{Analysis of the bottom-strange channel}
\label{sec:analysis-bs}

\begin{figure}[tp]
    \begin{tabular}{cc}
        \includegraphics*[height=0.45\textwidth]{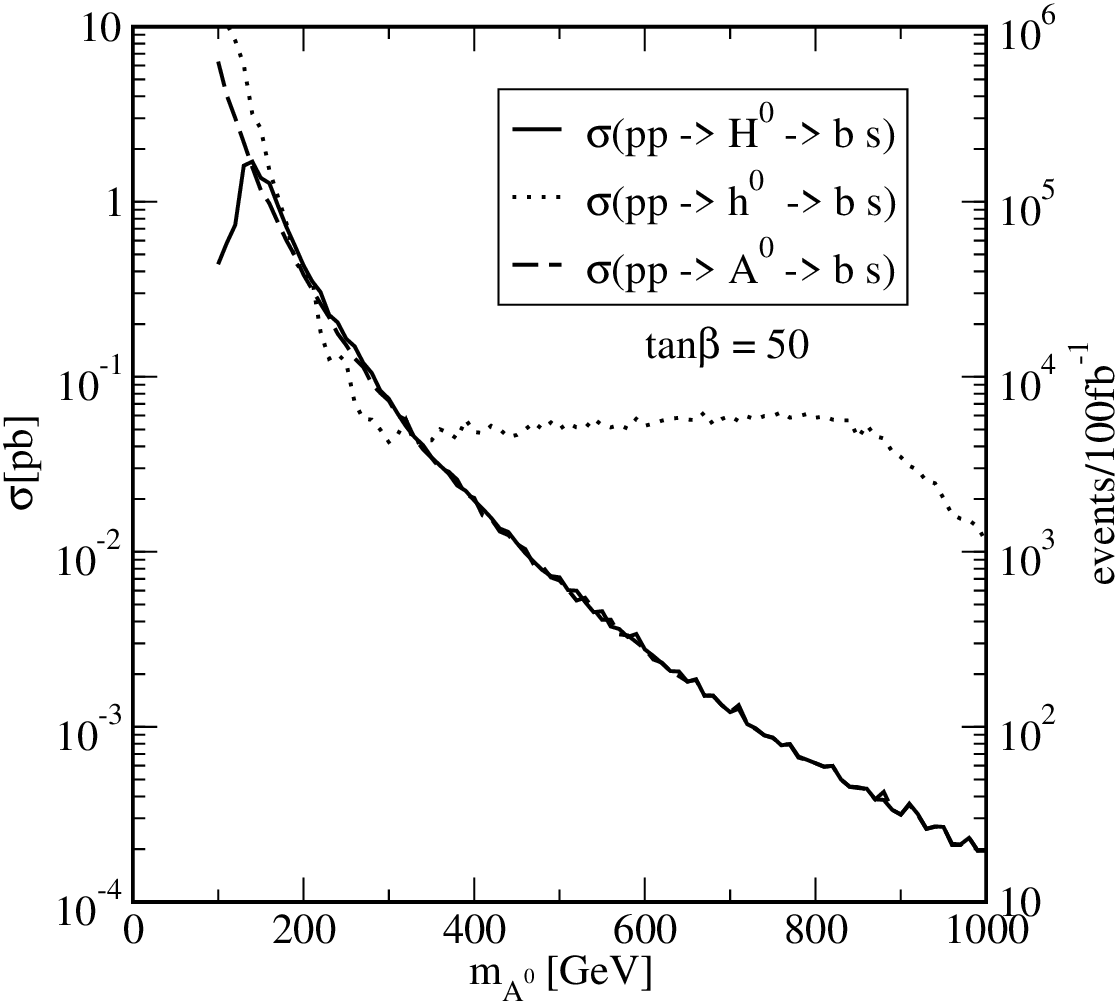} &
        \includegraphics*[height=0.45\textwidth]{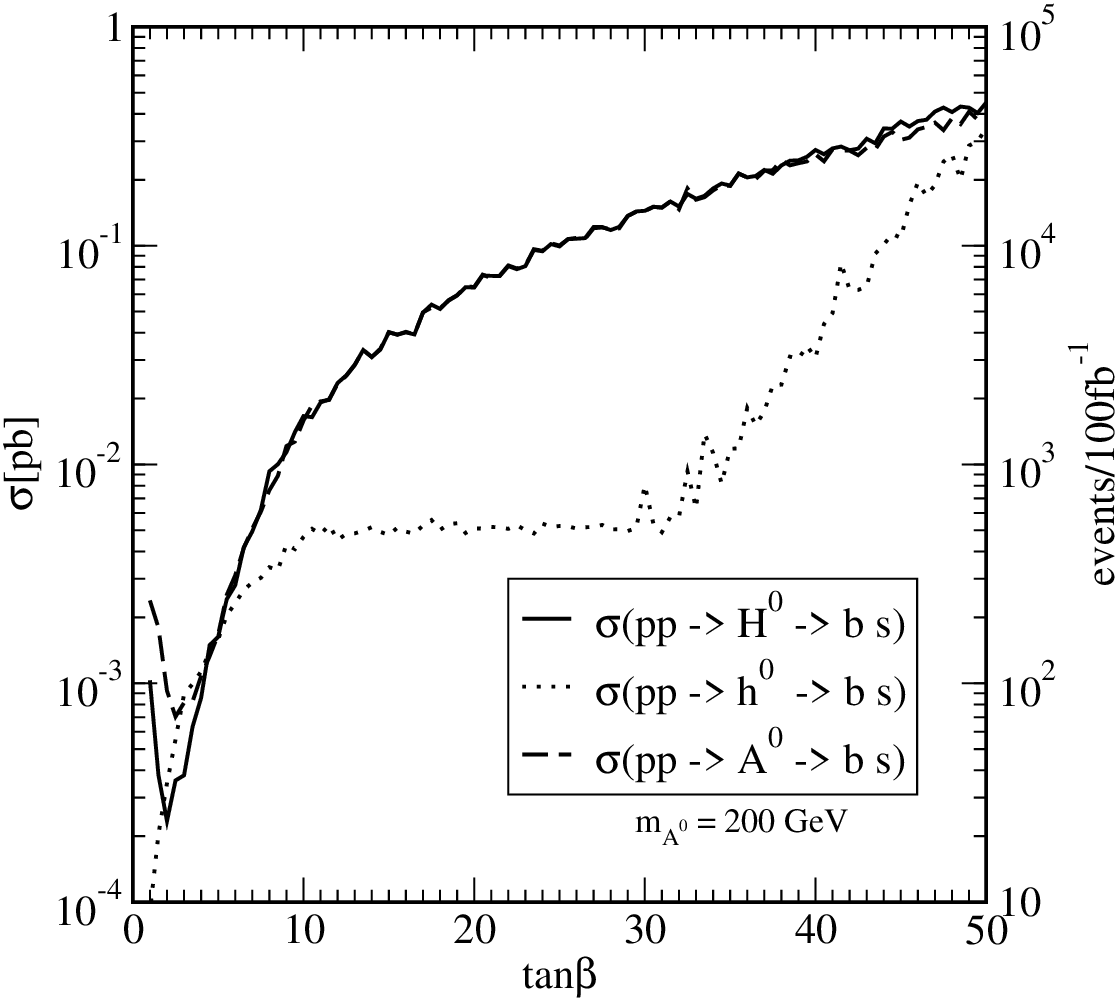} \\
        (a) & (b)
    \end{tabular}
    \caption{Maximum SUSY-QCD contributions to $\sigma(pp\rightarrow
      h\rightarrow b\,s)$, Eq.\,(\ref{eq:hqq-def}), as a function of
      \textbf{(a)} $\mA$ (at fixed $\tb$) and \textbf{(b)} $\tan\beta$ (at fixed $\mA$).
      In each plot the
      left-vertical axis provides the cross-section in pb and the
      right-vertical axis tracks the number of events per $100 \fb^{-1}$
      of integrated luminosity.}
    \label{fig:hbs-prod-ma-tb}
\end{figure}

The main result of the numerical scan for the bottom channel is
shown in Fig.~\ref{fig:hbs-prod-ma-tb}.  To be specific:
Fig.~\ref{fig:hbs-prod-ma-tb}a shows the maximum values of the
production cross-sections $\sigmapphbs$ for the three MSSM Higgs
bosons $h=h^0,H^0,A^0$ at the LHC, as a function of $\mA$;
Fig.~\ref{fig:hbs-prod-ma-tb}b displays the cross-section as a
function of $\tan\beta$. In this plot we indicate the value of
$\sigmapphbs$ (in pb) in the left-vertical axis, and at the same
time we track number of FCNC events (per $100 \fb^{-1}$ of
integrated luminosity) on the right vertical axis. Looking at
Fig.~\ref{fig:hbs-prod-ma-tb} one can see immediately that at
large $\tb$: i) the maximum number of events is remarkably high ($
10^6$ events!) for a FCNC process; ii) there is a sustained
region in the $h^0$ channel, comprising the interval
$300\GeV\lesssim\mA\lesssim900\GeV$, with a flat value of
$5\times 10^3$ events; iii) the chosen value of $\tan\beta=50$ is
not critical for $H^0$ and $A^0$ as long it is larger than $10$.
In Fig.~\ref{fig:hbs-prod-ma-tb}b  we see that the dependence on
$\tb$ is essentially the same for $H^0$ and $A^0$, but for $h^0$
it is quite different: in the region
$10\lesssim\tan\beta\lesssim30$ the cross-section remains below
$10^{-2}\pb$, but for $\tan\beta>30$ it starts climbing fast up
to $0.3\pb$ at $\tan\beta=50$. The number of events here reaches
a few times $10^4$ for all channels (for fixed
$\mA=200\GeV$)\,\footnote{As already advertised, in reading the plots
  and tables in this work, 
  one must keep in mind that they are the result of a Monte-Carlo sampling
  near the region of the maximal values.}.
For further reference, in Table~\ref{tab:hbs-maxims} we show the
numerical values of $\sigmapphbs$ together with the parameters
which maximize the production for $\tb=50$ and $\mA=200\GeV$.
We include the value of $B(h\to bs)$ at the maximization point of
the FCNC cross-section. We notice that at this point the
lightest Higgs boson $h^0$ is the one having the smallest branching
ratio. This is in contrast to the situation when one maximizes the
branching ratios independently of the number of events\,\cite{\BejarJHEP}.

\begin{figure}[tp]
    \begin{tabular}{cc}
        \includegraphics*[height=0.45\textwidth]{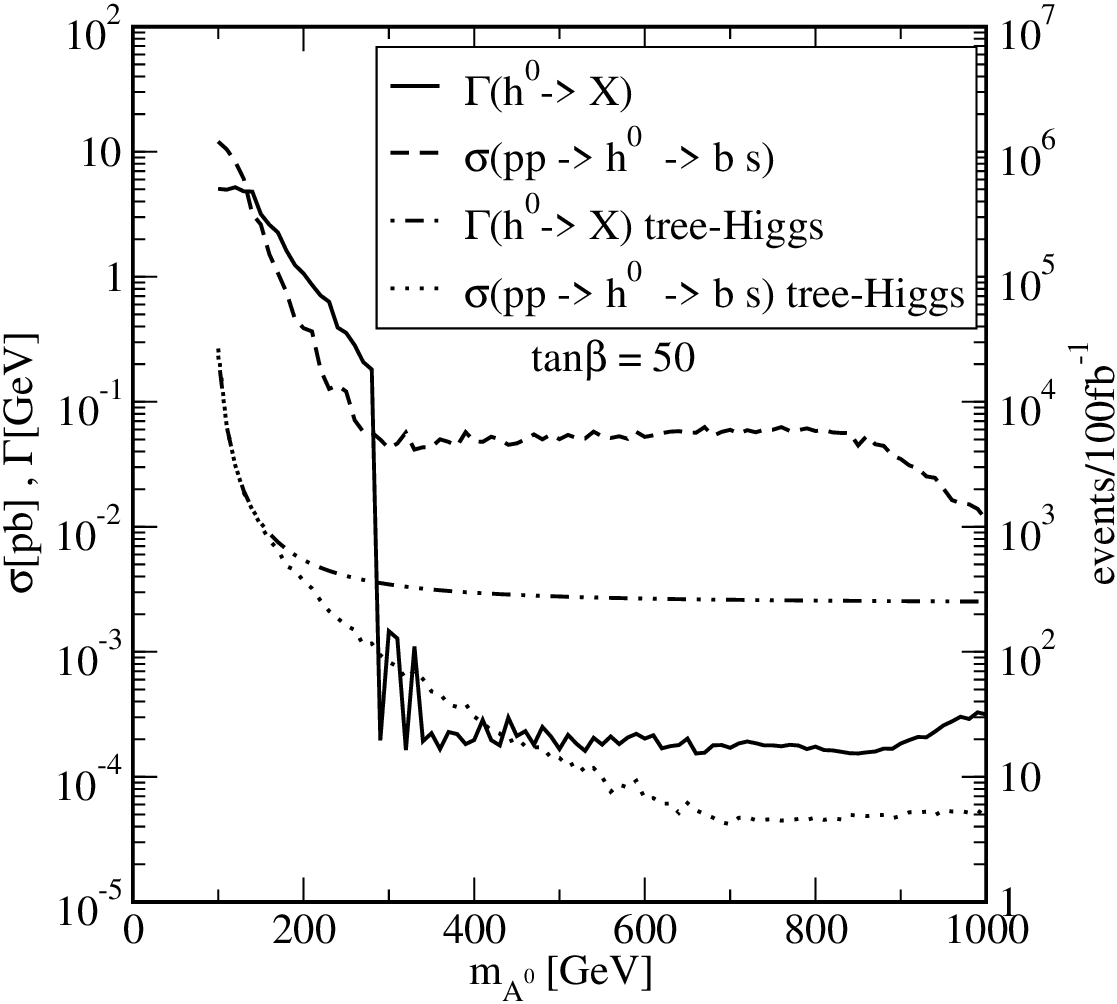} &
        \includegraphics*[height=0.45\textwidth]{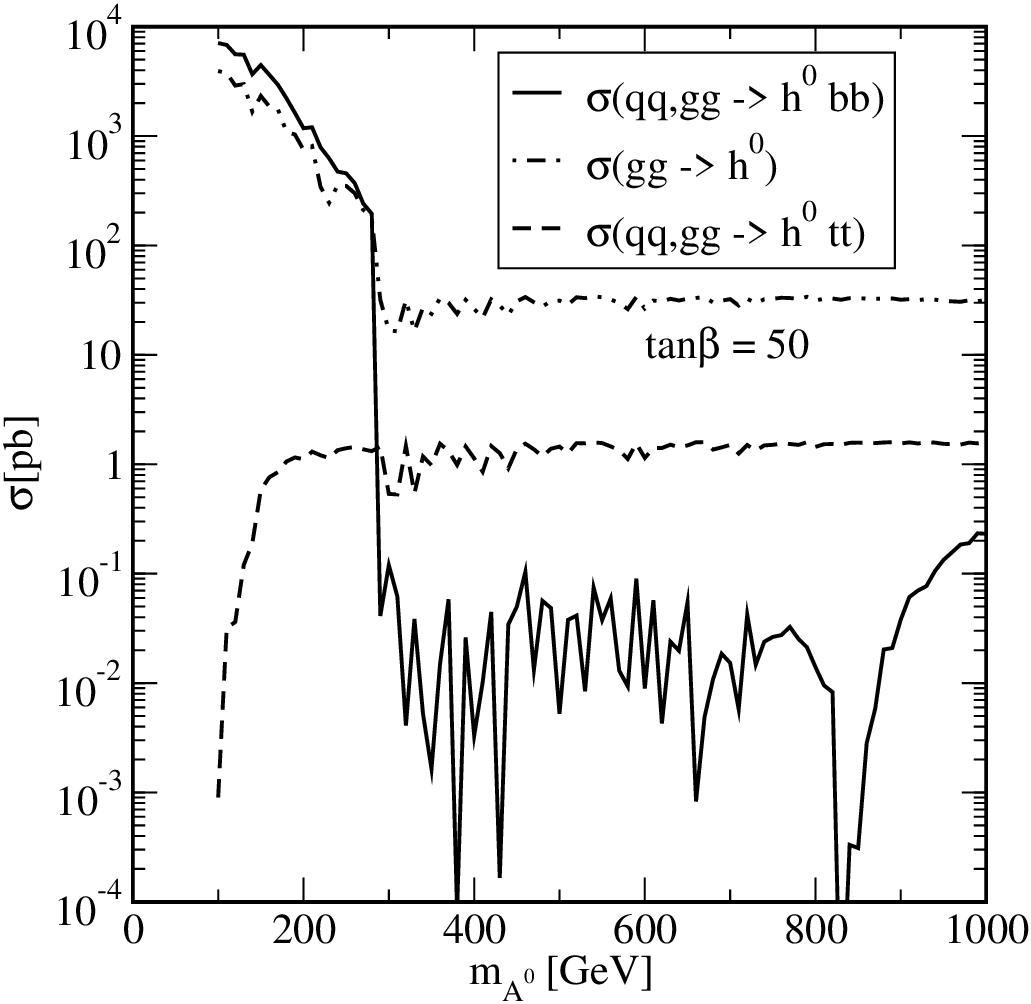} \\
        (a) & (b)
    \end{tabular}
    \caption{\textbf{(a)} $h^0$ production cross-section and decay width as a
      function of $\mA$ with the Higgs mass relations at tree-level
      and at
      one-loop. \textbf{(b)} Different contributions to the $h^0$
      production cross-section as a function of $\mA$
      corresponding to the maximization of $\sigmapphzbs$ using
      the one-loop Higgs mass relations. }
    \label{fig:hbs-prod-gamma}
\end{figure}

In Fig.~\ref{fig:hbs-prod-gamma}a we show the effect on
$\sigmapphzbs$, and on the total decay width
$\Gamma(h^0\rightarrow X)$, of using the Higgs boson sector at the
tree-level or at one-loop in our computation. It is well-known
that the $h^0$ couplings to quarks are particularly sensitive to
this issue, and for this reason we focus on the lightest CP-even
Higgs boson for these considerations. We can appreciate the
correlations among the different factors that enter the production
rate~(\ref{eq:hqq-def}). The plotted values for $\sigmapphzbs$
and $\Gamma(h^0\rightarrow X)$ in Fig.~\ref{fig:hbs-prod-gamma}a
correspond precisely to the parameters that
maximize~(\ref{eq:hqq-def}) at the tree-level or at one loop in
each case. Fig.~\ref{fig:hbs-prod-gamma}b shows a comparison of
the various $h^0$-production mechanisms for the values that
maximize $\sigmapphzbs$ at one-loop. Remarkably, the effect of the
radiative corrections in the Higgs sector amounts to an
enhancement of our maximal FCNC rates of up to three orders of
magnitude. As we mentioned in the introduction (see also
Ref.\cite{\BejarJHEP}) the maximum of $B(h^0\to b\bar{s})$ is
attained under the conditions of the so-called ``small
$\alpha_{\rm eff}$ scenario''\,\cite{Carena:2002qg,Carena:1999bh},
where the two-body decay $h^0\to b\bar{b}$ is strongly suppressed
due to a corresponding suppression of the $h^0\,b\,\bar{b}$
coupling. Since $\Gamma(h^0\to b\bar{b})$ usually dominates the
total width $\Gamma(h^0\to X)$, the latter also diminishes
drastically (at the level of the partial width of a $h^0$
three-body decay, as mentioned above). In this scenario the
production cross-section $\sigma(pp\to h^0 b\bar{b})$ is also
suppressed, so the final result is a compromise between the
suppression of $\Gamma(h^0\to b\bar{b})$ and the possible
enhancement (or at least sustenance) of $\sigma(pp\to h^0\,X)$ by
other mechanisms other than the associated production with bottom
quarks (like the mechanism of gluon-gluon fusion, see
Fig.~\ref{fig:hbs-prod-gamma}b).
\begin{table}
    \center
    \begin{tabular}{|c||c|c|c|}
        \hline
        $h$ &  $H^0$ & $h^0$ & $A^0$ \\\hline\hline
        \sigmapphbs &  $0.45\pb$ & $0.34\pb$ & $0.37\pb$ \\\hline
        events/$100\fb^{-1}$ & $4.5\times10^4$ & $3.4\times 10^4$ & $3.7\times10^4$\\\hline
        $B(h\to bs)$ & $9.3\times 10^{-4} $& $2.1\times 10^{-4} $& $8.9\times10^{-4} $ \\\hline
        $\Gamma(h\to X)$ & $10.9\GeV$ & $1.00\GeV$ & $11.3\GeV$
        \\\hline
        $\delta_{23}$ & $10^{-0.62}$ & $10^{-1.32}$ & $10^{-0.44}$ \\\hline
        $m_{\squark}$ & $990\GeV$ &  $670\GeV$ & $990\GeV$ \\\hline
        $A_b$ & $-2750\GeV$ & $-1960\GeV$ & $-2860\GeV$ \\\hline
        $\mu$ & $-720\GeV$ & $-990\GeV$ & $-460\GeV$ \\\hline
        \Bbsg & $4.50\times 10^{-4}$ & $4.47\times 10^{-4}$ & $4.39\times
        10^{-4}$ \\\hline
    \end{tabular}
    \caption{Maximum value of $\sigmapphbs$ (and of the number of
      $bs$ events per $100\,fb^{-1}$) in the LHC,  for $\mA=200\GeV$ and
      $\tan\beta=50$. Shown are also the corresponding values of the
      relevant branching ratio $B(h\to bs)$ and of the total width of
      the Higgs bosons, together with the values of the SUSY
      parameters. The last row includes $B(\bsg)$.}
    \label{tab:hbs-maxims}
\end{table}

Indeed, for $\mA\lesssim300\GeV$ we can check in
Fig.~\ref{fig:hbs-prod-gamma}b that the most relevant factor for
maximizing the FCNC cross-section is the enhancement of the $h^0$
production channel in association with bottom quarks,
$\sigma(pp\to h^0 b\bar{b})$. This channel operates through the
$b\bar{b}$-fusion vertex $b\bar{b}\rightarrow h^0$ and is highly
enhanced at large $\tb$. In the region $\mA\lesssim300\GeV$ stays
as the dominant mechanism for $h^0$ production, although one can
see that the alternate $gg$-fusion mechanism remains all the way
non-negligible. The corresponding effect on our FCNC cross-section
(\ref{eq:hqq-def}) is nevertheless not obvious because this same
parameter choice does also maximize the total width of $h^0$,
mainly through the enhancement of $\Gamma(h^0 \to b\,\bar{b})$.
There is a delicate interplay of various factors here. In
particular, in the region $\mA<300\GeV$ the maximized partial
width of the FCNC process $h^0\rightarrow bs$ (which is a
function of all the parameters in (\ref{eq:scan-parameters})) is
larger than in the region $\mA>300\GeV$, but at the same time the
total width becomes smaller in the latter region. Overall, the
result is that $\sigmapphzbs$ is larger in the former $\mA$ range
than in the latter. Furthermore, when we cross ahead the limit
$\mA\simeq 300\GeV$ the dominant $h^0$-production channel changes
turn: the associated $h^0$-production with bottom quarks falls
abruptly down (see explanations below) and the gluon-gluon fusion
mechanism takes over, so that in this range the $h^0$-production
cross-section becomes completely dominated by $gg\rightarrow
h^0$. We note that while this mechanism is not so efficient at
its maximum as the associated production, it has a virtue: it is
non-suppressed in the entire range of $\mA$. As a result, in the
region above $\mA\gtrsim 300\GeV$ a small value of the
$h^0\,b\,\bar{b}$ coupling enhances $B(h^0\to bs)$ while it does
not dramatically suppress the total cross-section $\sigma(pp\to
hX)$. For $\mA>300\GeV$ our sampling procedure finds the maximum
by selecting the points in parameter space corresponding to the
small $\alpha_{\rm eff}$ scenario, where $B(h^0\to bs)$ takes the
highest values and $\Gamma(h^0\to X)$ and $\sigma(pp\to h^0
b\bar{b})$ are strongly suppressed -- the latter staying below
the associated Higgs boson production with top quarks!\, In this
way the net Higgs boson cross-section $\sigma(pp\to h^0)$ is not
drastically reduced thanks to the sustenance provided by the
$gg$-fusion channel.
\begin{figure}[tp]
    \begin{tabular}{cc}
        \includegraphics*[height=0.45\textwidth]{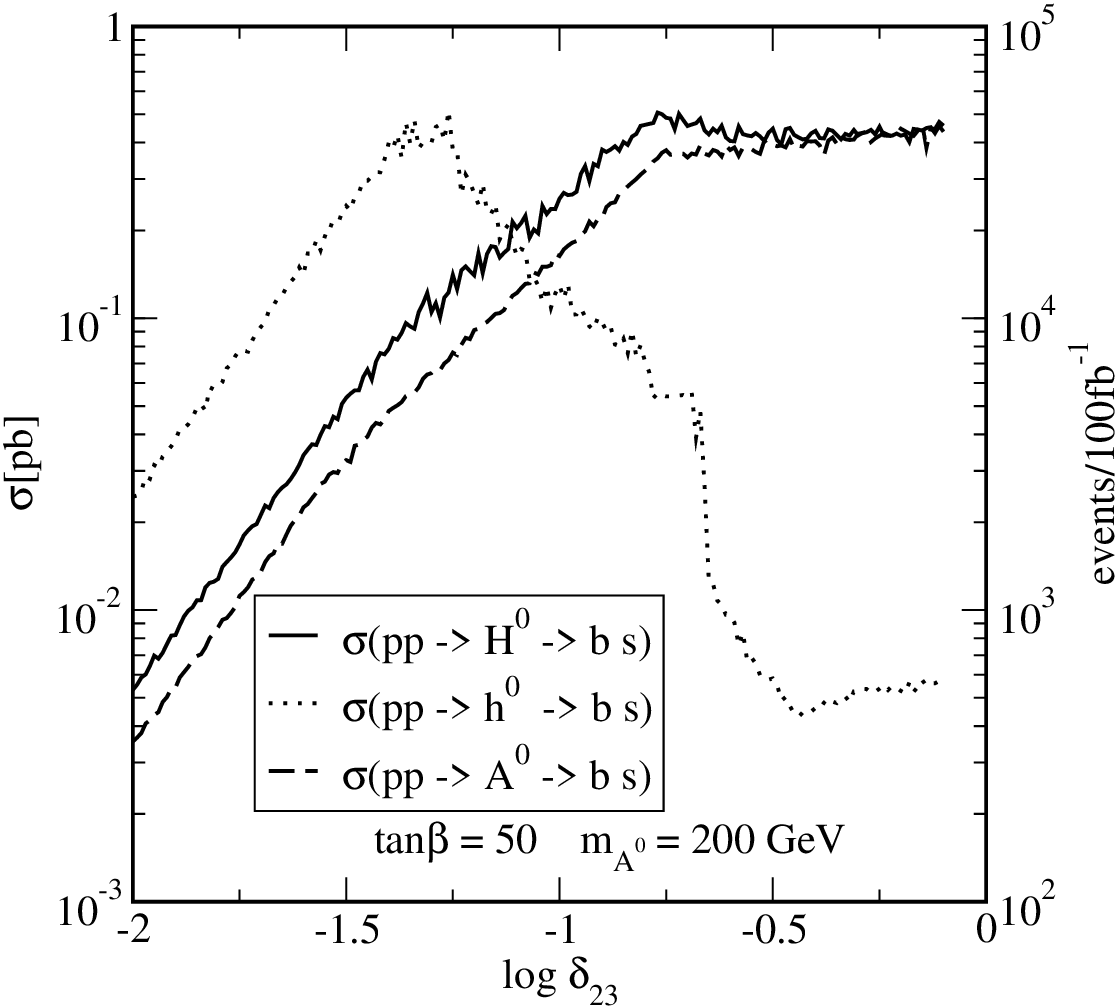} &
        \includegraphics*[height=0.45\textwidth]{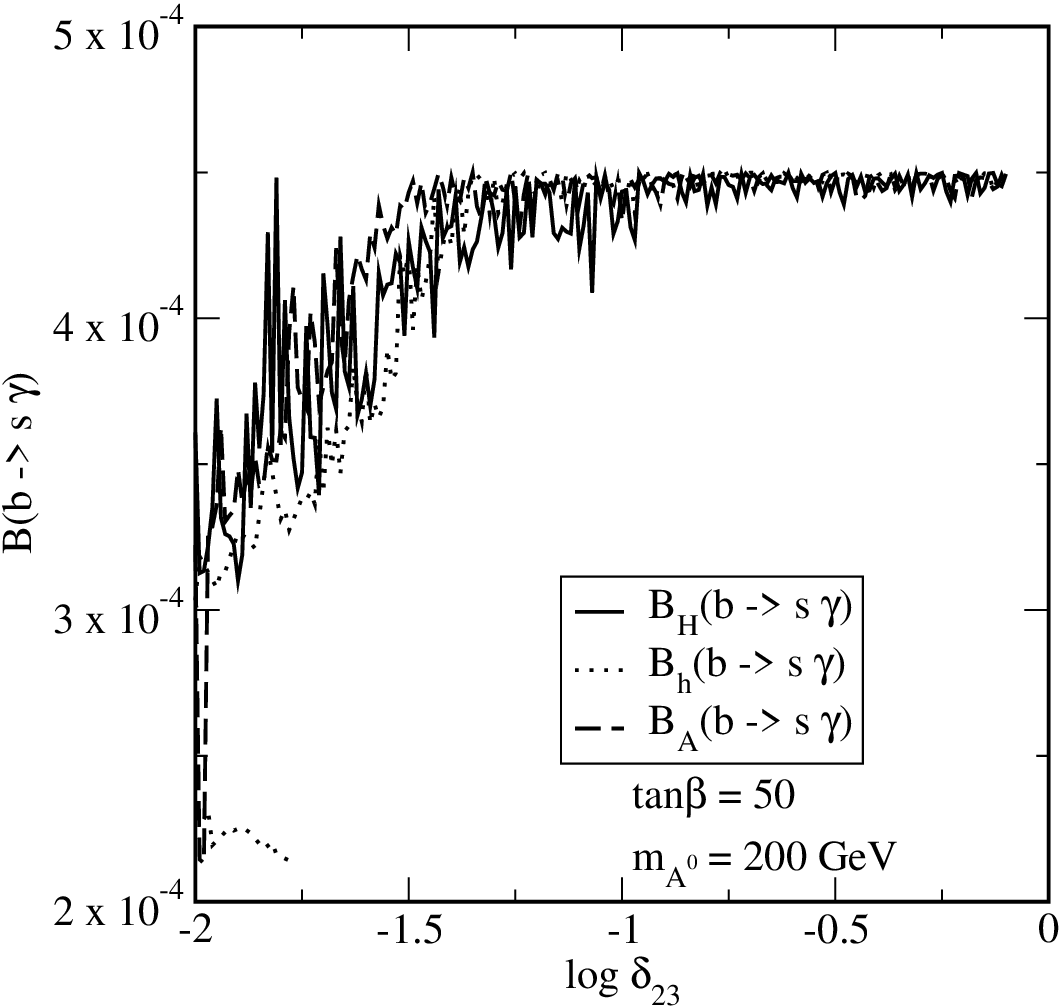} \\
        (a) & (b)
    \end{tabular}
    \caption{Maximum SUSY-QCD contributions to
      $\sigma(pp\rightarrow h\rightarrow b\,s)$, Eq.\,(\ref{eq:hqq-def}), as a
      function of \textbf{(a)}  $\deltatt$ and \textbf{(b)} value of $\bsg$.}
    \label{fig:hbs-prod-bsg-d23}
\end{figure}
Most of the loop contributions to it come from the top quark
because the bottom quark contribution is suppressed and the
squarks are rather heavy. One can clearly see this sustenance
feature in Fig.~\ref{fig:hbs-prod-gamma} in the form of a long
cross-section plateau up to around $\mA=1\TeV$, beyond which the
small $\alpha_{\rm eff}$ scenario cannot be maintained and
$\Gamma(h^0\to X)$ starts increasing and at the same time
$\sigmapphzbs$ starts decreasing. It is remarkable that this
behavior is only feasible thanks to the large radiative
corrections in the MSSM Higgs sector. When we, instead, perform
the computation using the tree-level relations for the Higgs
sector, the small $\alpha_{\rm eff}$ scenario is obviously not
possible and the enhancements/suppressions of $\sigma(pp\to h^0
b\bar{b})$/$\Gamma(h^0\to X)$ cannot take place. As a result the
FCNC rate is some 3 orders of magnitude smaller than in the
previous case.
\begin{figure}[tp]
    \begin{tabular}{cc}
        \includegraphics*[height=0.45\textwidth]{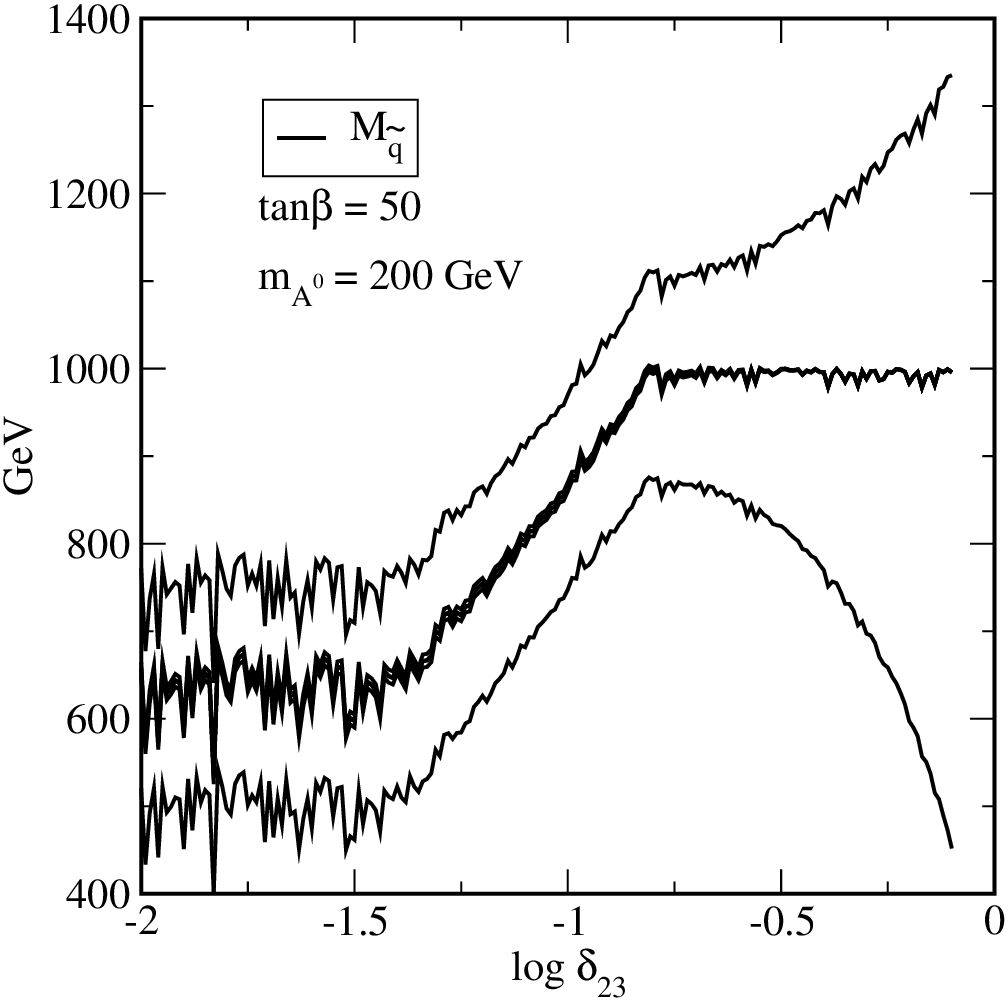} &
        \includegraphics*[height=0.45\textwidth]{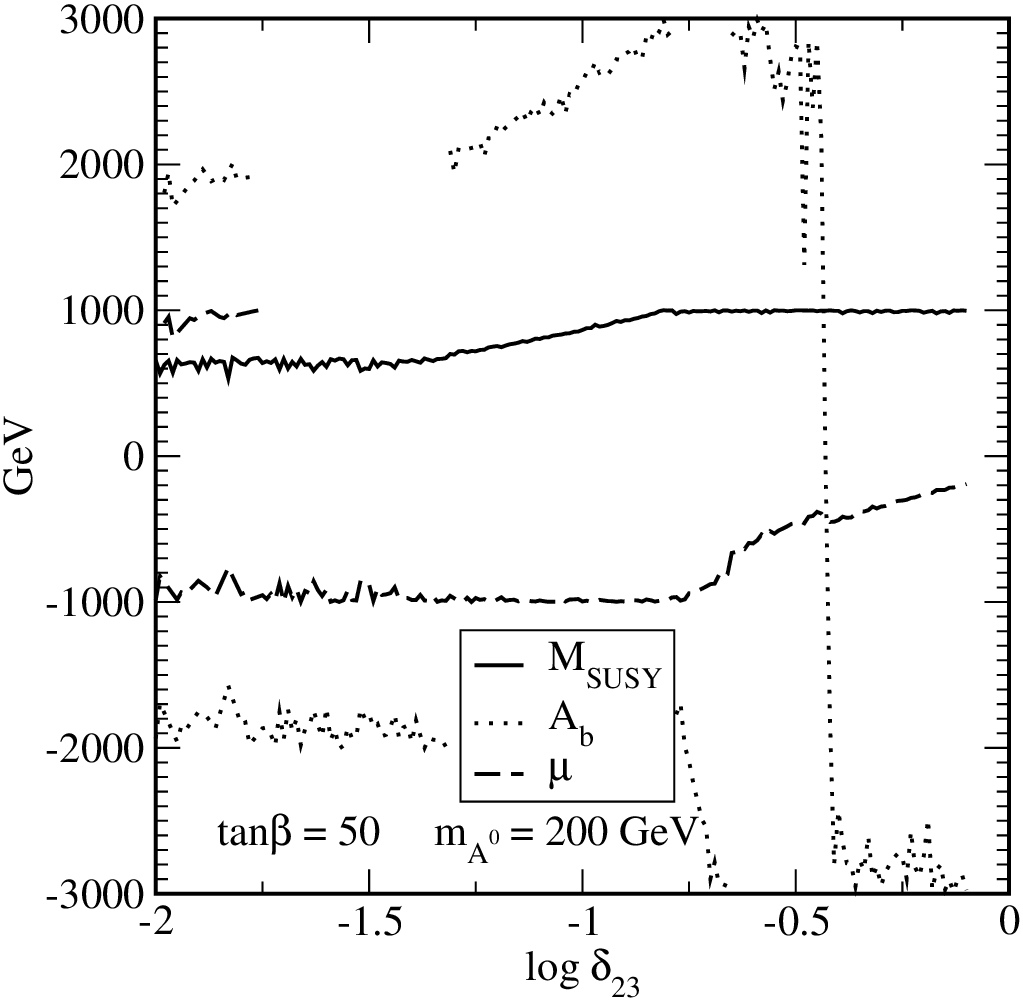} \\
        (a) & (b)
    \end{tabular}
    \caption{Value of \textbf{(a)} down-type squark physical  masses
      ($M_{\tilde{q}}$), four of them are degenerate; \textbf{(b)}
      the
      parameters~(\ref{eq:scan-parameters}) from the maximization of
      the $h^0$ channel in Fig.~\ref{fig:hbs-prod-bsg-d23}a.}
    \label{fig:hbs-msq-param-d23}
\end{figure}

Next, we turn our attention to the FCNC mixing parameter
$\deltatt$ in Eq.~(\ref{delta23}). The value of \deltatt\ at the
cross-section maximum is not necessarily the maximum allowed
value of \deltatt\ in (\ref{eq:scan-parameters}). This is because
it is a conditioned maximum, namely a maximum obtained under the
restrictions imposed by $\bsg$, as illustrated in
Fig.~\ref{fig:hbs-prod-bsg-d23}. For further reference in
our discussion, and to better grasp some qualitative features of
our results, let us write the general form of the SUSY-QCD
contribution to $\bsg$. If we emphasize only the relevant
supersymmetric terms under consideration (obviating the powers of
the gauge couplings and other factors) we have
\begin{equation}\label{bsgamma}
B(\bsg)\sim \deltatt^2\,\frac{m_b^2(\mu-A_b\tb)^2}{\msusy^4}\,.
\end{equation}
Fig.~\ref{fig:hbs-prod-bsg-d23}a shows the maximum of
$\sigmapphbs$ as a function of $\deltatt$ for a fixed value of
the CP-odd Higgs boson mass $\mA=200\GeV$, whereas
Fig.~\ref{fig:hbs-prod-bsg-d23}b shows the computed value of
$\Bbsg$ corresponding to the parameter space points where each
maximum is attained. At small $\deltatt$ the SUSY-QCD
contribution to \Bbsg\ is negligible, and the experimental
restriction $\Bbsg=(2.1-4.5)\times 10^{-4}$ does not place
constraints on the other MSSM
parameters~(\ref{eq:scan-parameters}); in other words, in this
region the dependence is $\sigmapphbs\propto (\deltatt)^2$ -- the
naively expected one. Here $\deltatt\lesssim 10^{-1.5}\simeq
0.03$ and the computed $\Bbsg$ value lies well within the
experimental limit. For larger $\deltatt$, $\Bbsg$ can be
saturated at its uppermost experimentally allowed limit, and the
rest of the parameters in~(\ref{eq:scan-parameters}) must change
accordingly in order not to cross that limit. This can be
appreciated in Fig.~\ref{fig:hbs-msq-param-d23} where we show the
range of values taken by the physical down-type squark
masses\footnote{There are six different down-type squarks, but
four of them are nearly degenerate in mass in our approximation.
For the $bs$-channel, the down squarks are so heavy that the
conditions required in the last two rows of
(\ref{eq:scan-parameters}) are automatically satisfied by them in
practically all the allowed range for $\mA$. }
(Fig.~\ref{fig:hbs-msq-param-d23}a) and the lagrangian
parameters~(Fig.~\ref{fig:hbs-msq-param-d23}b) from
Eq.~(\ref{eq:scan-parameters}) that provide the maximum values of
$\sigmapphzbs$ in Fig.~\ref{fig:hbs-prod-bsg-d23}. In the small
$\deltatt$ region ($\deltatt\lesssim10^{-1.5}\simeq 0.03$) the
parameters and masses that maximize $\sigmapphbs$ are constant --
except for the statistical noise unavoidable in a Monte-Carlo
procedure. Note also that there are two possible values for the
parameters $\mu$ and $A_b$, due to the fact that (the leading
contribution of) $\sigmapphbs$ is independent of the sign of
these parameters and the Monte-Carlo procedure picks either sign
for each point with equal probability. At $\deltatt\simeq
10^{-1.5}$ the value of $\Bbsg$ becomes saturated and
$\sigmapphzbs$ reaches its maximum (cf.
Fig.~\ref{fig:hbs-prod-bsg-d23}b); however $\deltatt$ can keep
growing, yet without overshooting the $\Bbsg$ limits, because the
increasing value of $\deltatt$ is compensated by the growing
squark masses (cf. Fig.~\ref{fig:hbs-msq-param-d23}a). But this
is not all that simple, the higher range of $\deltatt$ can be
further divided in two more segments where different dynamical
features occur.  In the first range, namely $10^{-1.5} \lesssim
\deltatt \lesssim 10^{-0.75}$, the heavy Higgs boson channels
keep on increasing their FCNC rates, but not so the lightest
Higgs boson channel $\sigmapphzbs$, the reason being that for
higher squark masses we reach the region where the small
$\alpha_{\rm eff}$ scenario is feasible and hence the $h^0$
couplings become weakened.  The relevant terms of the
cross-section can roughly be written as follows (see
Ref.\cite{\BejarJHEP}):
\begin{equation}\label{sigmahz}
\sigmapphzbs\sim\sigma(pp\rightarrow h^0)\times \deltatt^2
\cos^2(\beta-\alpha_{\rm eff})\, \mg^2\,\mu^2/\msusy^4\,,
\end{equation}
so that for large $\tb$ and small $\alpha_{\rm eff}$ it becomes
reduced. In the second high range of $\deltatt$, i.e. for
$\deltatt\gtrsim 10^{-0.75}\simeq 0.18$, the SUSY mass parameter
$\msusy$ has already reached its allowed maximum value specified
in~(\ref{eq:scan-parameters}), therefore other parameters have to
change to compensate for the larger $\deltatt$.
\begin{figure}[tp]
    \centering
    \includegraphics*[height=0.45\textwidth]{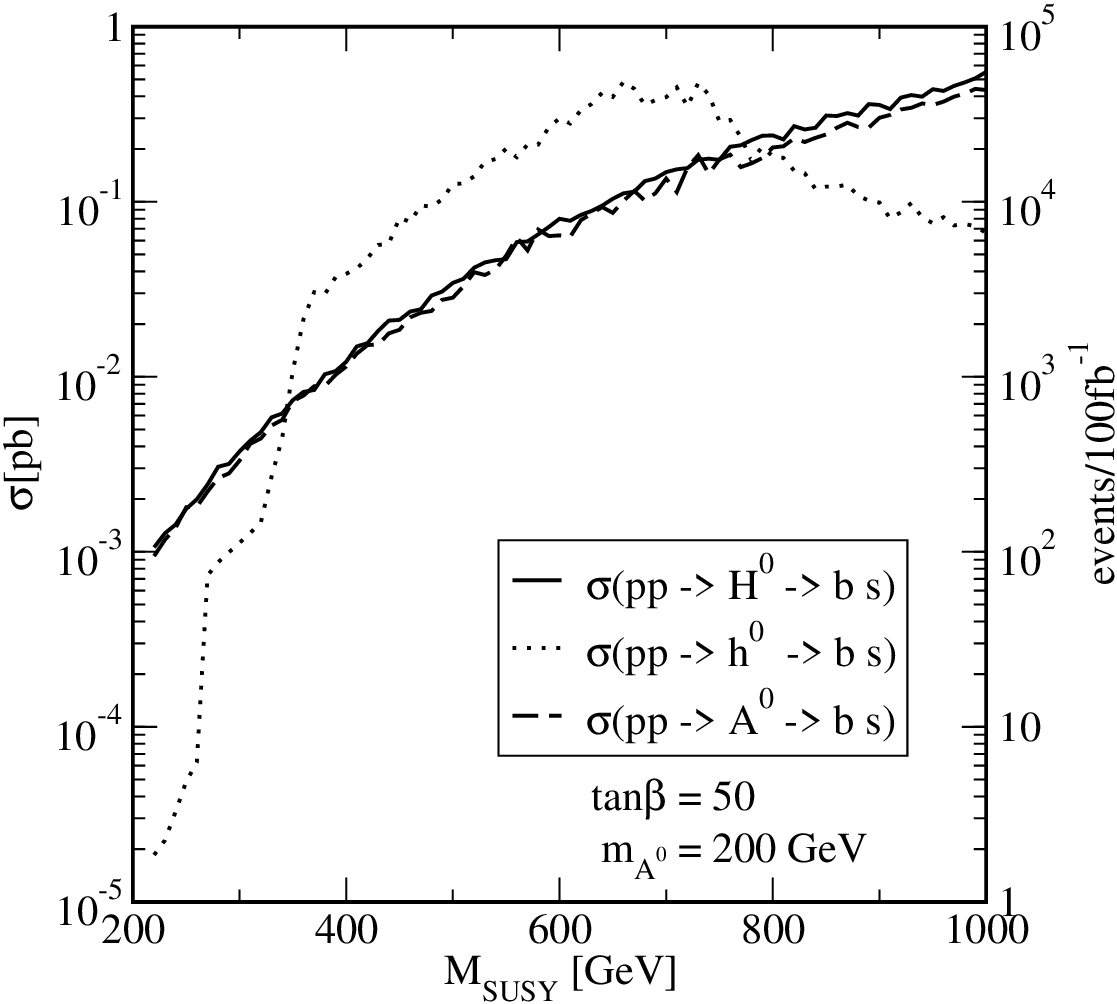}
    \caption{Maximum SUSY-QCD contributions to
      $\sigma(pp\rightarrow h\rightarrow b\,s)$, Eq.\,(\ref{eq:hqq-def}), as a
      function of \msusy.}
    \label{fig:hbs-prod-msusy}
\end{figure}
This is confirmed in Fig.~\ref{fig:hbs-msq-param-d23}b, where for
$\deltatt\gtrsim 10^{-0.75}$ the absolute value of $\mu$
decreases to preserve the $\Bbsg$ upper bound. Correspondingly,
in this region $\sigmapphzbs$ further falls down, as it is patent
in Fig\,\ref{fig:hbs-prod-bsg-d23}a. This additional feature can
also be understood from the approximate expression of the
cross-section given above. At the same time the FCNC rates for
$h=H^0,A^0$ keep further growing, but at a much lower pace. This
is because their (approximate) contribution goes like
\begin{equation}\label{sHA}
\sigma(pp\to (H^0,A^0)\to bs)\sim\sigma(pp\rightarrow
H^0,A^0)\times \deltatt^2\, (\sin^2(\beta-\alpha_{\rm
eff}),\,1)\,\mg^2\,\mu^2/\msusy^4\,,
\end{equation}
similar to the $h^0$ case but with angular dependences on
$\alpha_{\rm eff}$ and $\beta$ which are non-suppressing in this
region (see Ref.\cite{\BejarJHEP})\,\footnote{In Eq.(\ref{sHA})
above we have corrected a typing mistake present in Eq.\,(3.5) of
Ref.\,\cite{Bejar:2004rz}. }. Again, any further increase of
$\deltatt$ is now partially cancelled by the $\bsg$ constraint,
which demands smaller values of $\mu$. This explains the
stabilization of the FCNC rates of the $H^0$ and $A^0$ channels
in the highest $\deltatt$ range (cf.
Fig.~\ref{fig:hbs-prod-bsg-d23}a). The profile of the squark mass
curves in Fig.~\ref{fig:hbs-msq-param-d23}a implies a mixing mass
matrix with constant diagonal terms (with value $\msusy$) and
growing mixing terms ($\deltatt$).

We finish our analysis of $\sigmapphbs$ by looking at its
behavior as a function of the SUSY mass scale $\msusy$, viz. the
overall scale for the squark and gluino masses -- cf.
Eq.~(\ref{eq:scan-parameters}). Fig.~\ref{fig:hbs-prod-msusy}
shows the maximum of $\sigmapphbs$ as a function of $\msusy$ for
fixed $\mA=200\GeV$ and $\tb=50$. The interpretation of this
figure follows closely the results of the previous ones. At small
values of $\msusy$ the potentially large contribution to $\Bbsg$
has to be compensated  -- see Eq.\,(\ref{bsgamma}) -- by
small values of $\deltatt$ and/or $|\mu|$, resulting in a
(relatively) small value of $\sigmapphbs$. As $\msusy$ grows,
$\deltatt$ and $|\mu|$ can take larger values without disturbing
the restrictions from \Bbsg. The leading contribution to our FCNC
cross-sections is actually independent of the overall SUSY mass
scale $\msusy$, because for all Higgs boson channels we have
found the general behavior (leaving aside other terms mentioned
above)
\begin{equation}\label{generalsigma}
\sigmapphbs\sim\sigma(pp\rightarrow h)\times
\deltatt^2\,\mg^2\,\mu^2/\msusy^4\ \ \  (h=h^0,\,H^0,\,A^0)\,.
\end{equation}
The last factor effectively behaves as
$\deltatt^2\,\mu^2/\msusy^2$, and grows as $\deltatt^2$ for
increasing $\msusy$ at fixed ratio $\mu/\msusy$. Under the same
conditions \Bbsg\ causes no problem because it is additionally
suppressed by $m_b^2/\msusy^2$ -- cf.
Eq.\,(\ref{bsgamma}). Therefore, we are led to a sort of
``non-decoupling behavior'' of the FCNC rates with increasing
$\msusy$. In other words, we find that for the heavy neutral
Higgs bosons ($H^0,A^0$) the interesting region is (contrary to
naive expectations) the high $\msusy$ range!\, The lightest Higgs
boson ($h^0$) channel shows a similar overall behavior, but it
presents additional features because it is more tied to the
evolution of the CP-even mixing angle $\alpha$. The most
interesting region for this channel is (cf.
Fig.~\ref{fig:hbs-prod-msusy}) the central squark mass scale
$\msusy\sim 600-800\GeV$, where the small $\alpha_{\rm eff}$
scenario can take place.

From the combined analysis of
Figs.~\ref{fig:hbs-prod-ma-tb}-\ref{fig:hbs-prod-msusy} we arrive
at the following conclusions concerning the $bs$ final state:
\begin{itemize}
\item A significant event rate of FCNC Higgs boson decays $\sigmapphbs$
  is expected at the LHC, even after taking into account the limits
  on $\Bbsg$;
\item Lightest Higgs boson case, $h^0$:
  \begin{itemize}
  \item  For $\mA\lesssim300\GeV$ the rate $\sigmapphzbs$ decreases
    with $\mA$ but it is the largest in this interval, being
    produced by the combination of a large production cross-section
    $\sigma(pp\to h^0 b\bar{b})$ and a moderate $B(\hzbs)$. It
    amounts to a number of events between $\sim5\times 10^3$ and
    $\sim 12\times 10^5$ for every $100\fb^{-1}$ of integrated luminosity
    at the LHC;
  \item For $300\GeV\lesssim\mA\lesssim 850\GeV$ we expect a maximum
    of $\sim 6 \times 10^3$ events/$100\fb^{-1}$ in the small $\alpha_{\rm eff}$ scenario,
    provided by a large $B(\hzbs)$ and $\sigma(pp(gg)\to h^0)$ as the
    dominant production cross-section;
  \item For $\mA>850\GeV$ the number of events starts to decrease
    slowly;
  \item In all cases, this maximum is attained for a large value of
    $\tb\sim50$, a moderate value of
    the SUSY mass scale ($\msusy\sim 600-800\GeV$) and a \textit{low}
    value of $\deltatt\sim10^{-1.3}\sim0.05$;
  \end{itemize}
\item Heavy Higgs bosons, $H^0, A^0$:
  \begin{itemize}
  \item Although not shown in our plots, we have checked that their production
  rate $\sigma(pp\to H^0\,A^0)$ decreases fast
  with the Higgs boson mass
    (due to the decreasing of the production cross-section).
    We find a maximum FCNC rate of $\sim5\times 10^4$ events for
    $\mA\simeq 200\GeV$, and $20$ events for $\mA\simeq 1\TeV$.
  \item The maximum is produced at i) large $\tb>30$, ii) at the
  highest allowed values of the SUSY mass scale, $\msusy\sim1\TeV$,
  and iii) at a relatively large value of the FCNC mixing
  parameter, $\deltatt\sim 10^{-0.75}\sim0.18$, but not at the largest allowed
    value. The small $\alpha_{\rm eff}$ scenario plays no role in the heavy
    Higgs boson channels.
  \end{itemize}
\end{itemize}
Altogether one should expect a total maximum of some $120,000$
events/$100\fb^{-1}$.

\section{Analysis of the top-charm channel}
\label{sec:analysis-tc}

\begin{figure}[tp]
    \begin{tabular}{cc}
        \includegraphics*[height=0.45\textwidth]{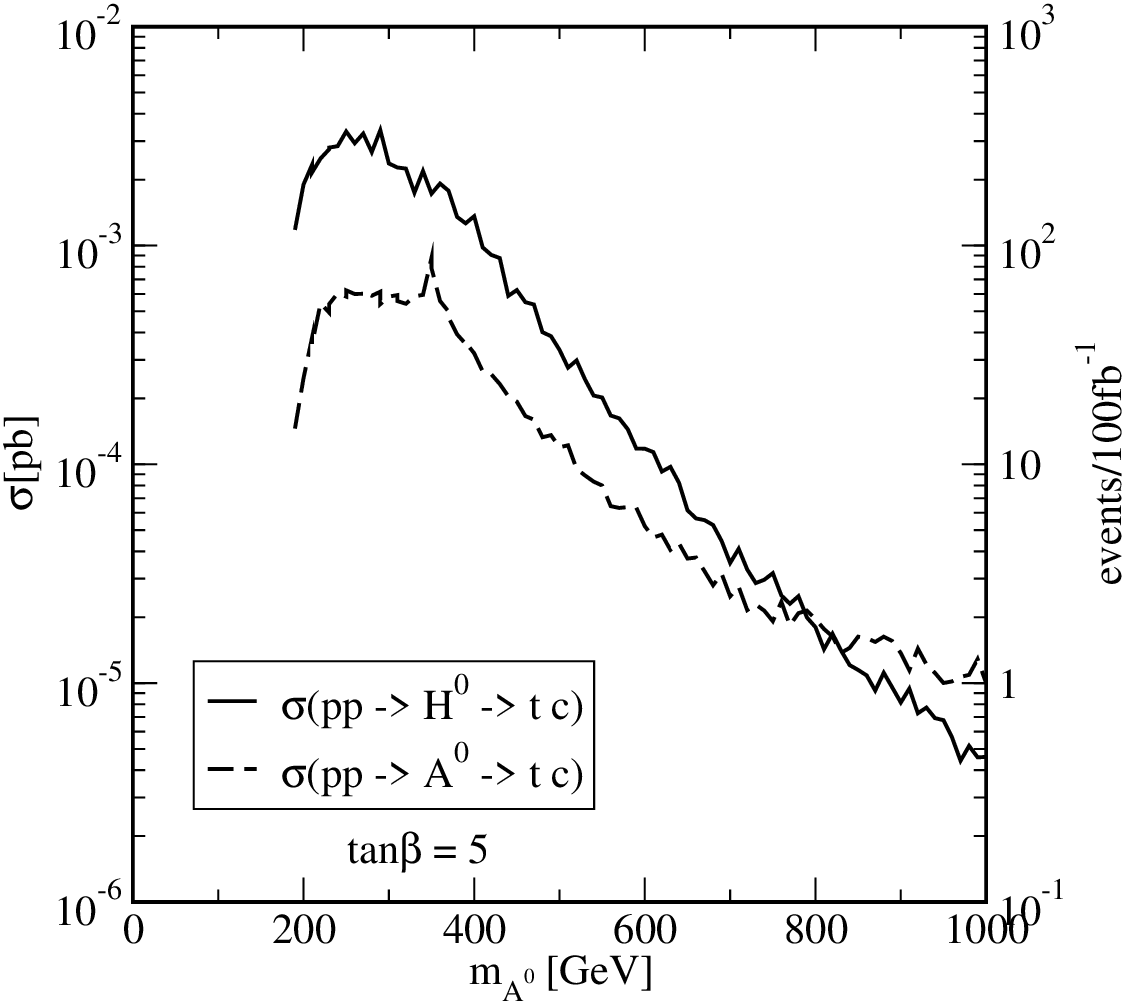} &
        \includegraphics*[height=0.45\textwidth]{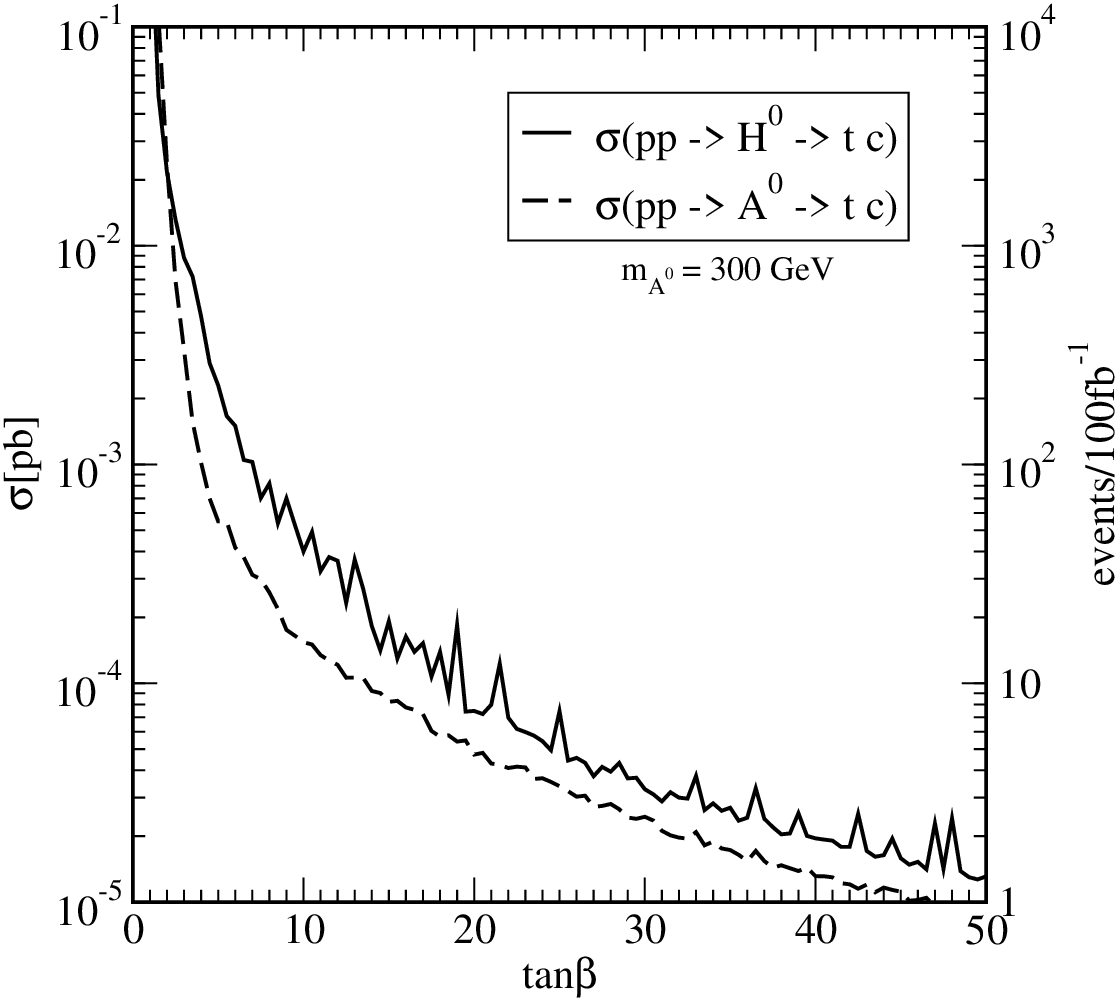} \\
        (a) & (b)
    \end{tabular}
    \caption{Maximum SUSY-QCD contributions to $\sigma(pp\rightarrow
      h\rightarrow t\,c)$, Eq.\,(\ref{eq:hqq-def}), as a function of
      \textbf{(a)} $\mA$ (at fixed $\tb$) and \textbf{(b)} $\tan\beta$ (at fixed $\mA$).}
    \label{fig:htc-prod-ma-tb}
\end{figure}

The results of the numerical scan for this channel are similar to
the $bs$ channel, so we will focus mainly on the differences.
Fig.~\ref{fig:htc-prod-ma-tb}a shows the maximum value of the
production cross-section $\sigmapphtc$, Eq.\,(\ref{eq:hqq-def}),
under study as a function of $\mA$; Fig.~\ref{fig:htc-prod-ma-tb}b
displays the cross-section as a function of $\tan\beta$.
Obviously the lightest Higgs boson ($h^0$) channel does not
appear in these plots, since in the MSSM this boson is always
lighter than the top quark.
 Looking at Fig.~\ref{fig:htc-prod-ma-tb} one can
see immediately the following: i) the dominant channel in this
case is the heavy scalar Higgs boson, $H^0$; ii) it varies between
$1$ and $300$ events/$100\fb^{-1}$; iii) the $\tan\beta$ value is
critical with preference for low values. In Table~\ref{tab:htc-maxims} we show
the numerical values of $\sigmapphtc$ together with the parameters which
maximize the production for $\tb=5$ and $\mA=300\GeV$.  We have included the
value of $B(h\to tc)$ at the maximization point of the FCNC cross-section. It
is remarkable that for the heavy CP-even Higgs boson one can reach $B(H^0\to
tc)\sim 10^{-3}$ compatible with the $\bsg$ constraint.

Let us remark that for the heavy Higgs boson channels the features
of the small $\alpha_{\rm eff}$ scenario play no significant role
because the partial widths into $b\bar{b}$ are proportional
either to $\cos^2\alpha_{\rm eff}$ (in the $H^0$ case) or to
$\sin^2\beta$ ($A^0$ case). Moreover, in the low $\tb\gtrsim 1$
region (the relevant allowed one for the $tc$ channel) the small
$\alpha_{\rm eff}$ scenario does not even have a chance to take
place.
\begin{table}
    \center
    \begin{tabular}{|c||c|c|c|}
        \hline
        $h$ &  $H^0$ & $A^0$ \\\hline\hline
        \sigmapphtc &  $2.4\times 10^{-3}\pb$ & $5.8\times 10^{-4}\pb$ \\\hline
        events/$100\fb^{-1}$ & 240 & 58  \\\hline
        $B(h\to tc)$ & $1.9\times 10^{-3} $& $5.7\times 10^{-4} $\\\hline
        $\Gamma(h\to X)$ & $0.41\GeV$ & $0.39\GeV$ \\\hline
        $\delta_{23}$ & $10^{-0.10}$ & $10^{-0.13}$ \\\hline
        $m_{\squark}$ & $880\GeV$ & $850\GeV$ \\\hline
        $A_t$ & $-2590\GeV$ & $2410\GeV$ \\\hline
        $\mu$ & $-700\GeV$ & $-930\GeV$ \\\hline
        \Bbsg & $4.13\times 10^{-4}$ & $4.47\times 10^{-4}$ \\\hline
    \end{tabular}
    \caption{Maximum value of $\sigmapphtc$ (and of the number of
      $tc$ events per $100\,fb^{-1}$) in the LHC,  for $\mA=300\GeV$ and
      $\tan\beta=5$. Shown are also the corresponding values of the
      relevant branching ratio $B(h\to tc)$ and of the total width of
      the Higgs bosons, together with the values of the SUSY
      parameters. The last row includes $B(\bsg)$.}
    \label{tab:htc-maxims}
\end{table}

\begin{figure}[tp]
    \begin{tabular}{cc}
        \includegraphics*[height=0.45\textwidth]{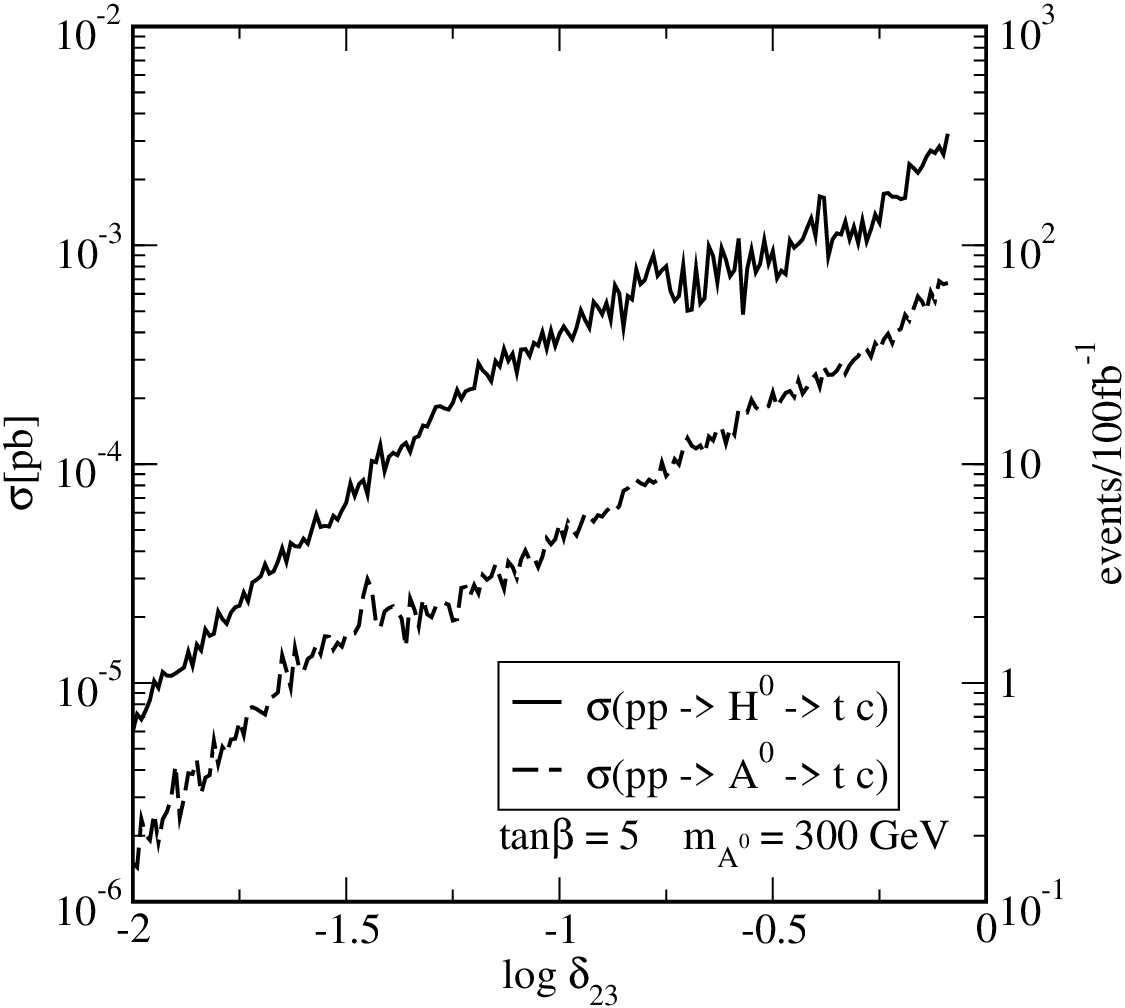} &
        \includegraphics*[height=0.45\textwidth]{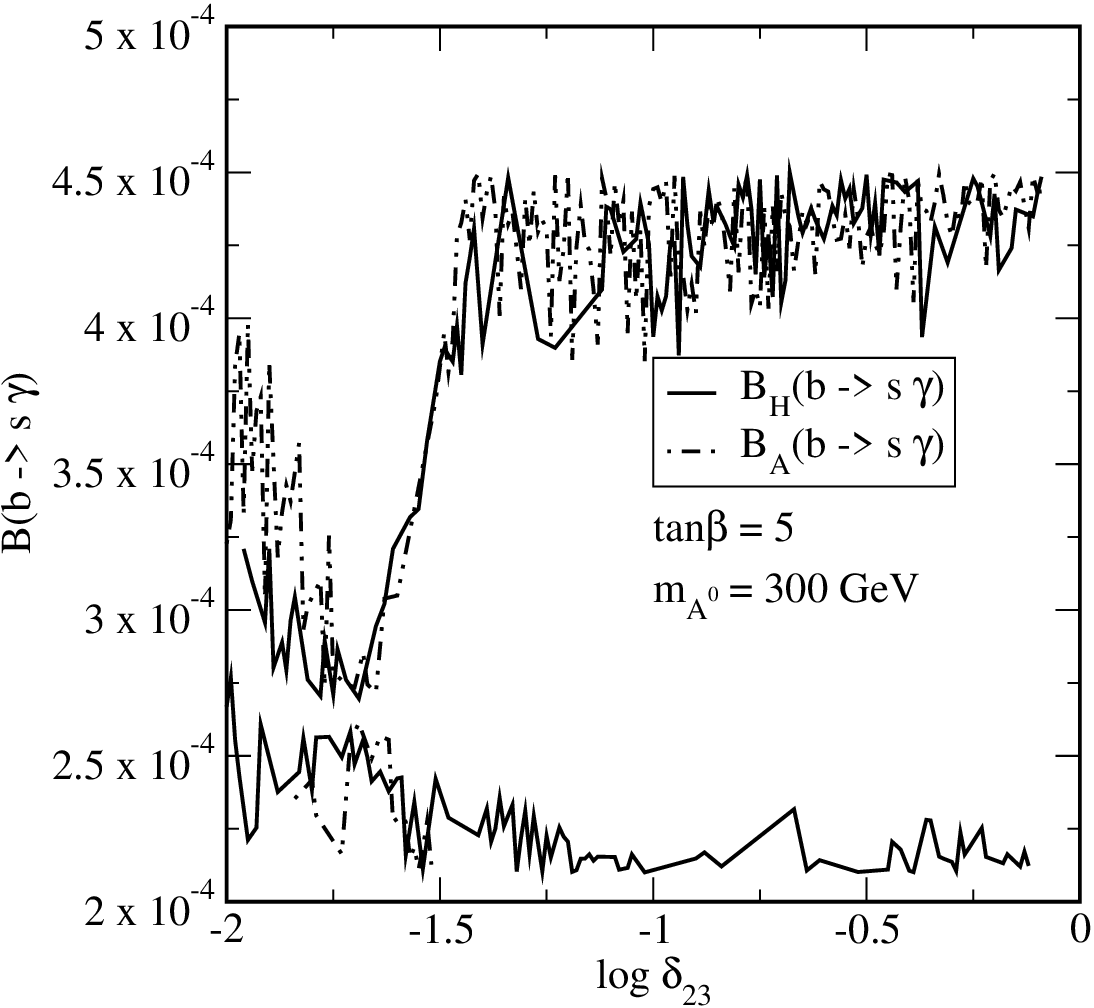} \\
        (a) & (b)
    \end{tabular}
    \caption{Maximum SUSY-QCD contributions to
      $\sigma(pp\rightarrow h\rightarrow t\,c)$, Eq.\,(\ref{eq:hqq-def}), as a
      function of \textbf{(a)} $\deltatt$ and \textbf{(b)} value of $\bsg$, for fixed $\tb$ and $\mA$.}
    \label{fig:htc-prod-bsg-d23}
\end{figure}

\begin{figure}[tp]
    \begin{tabular}{cc}
        \includegraphics*[height=0.45\textwidth]{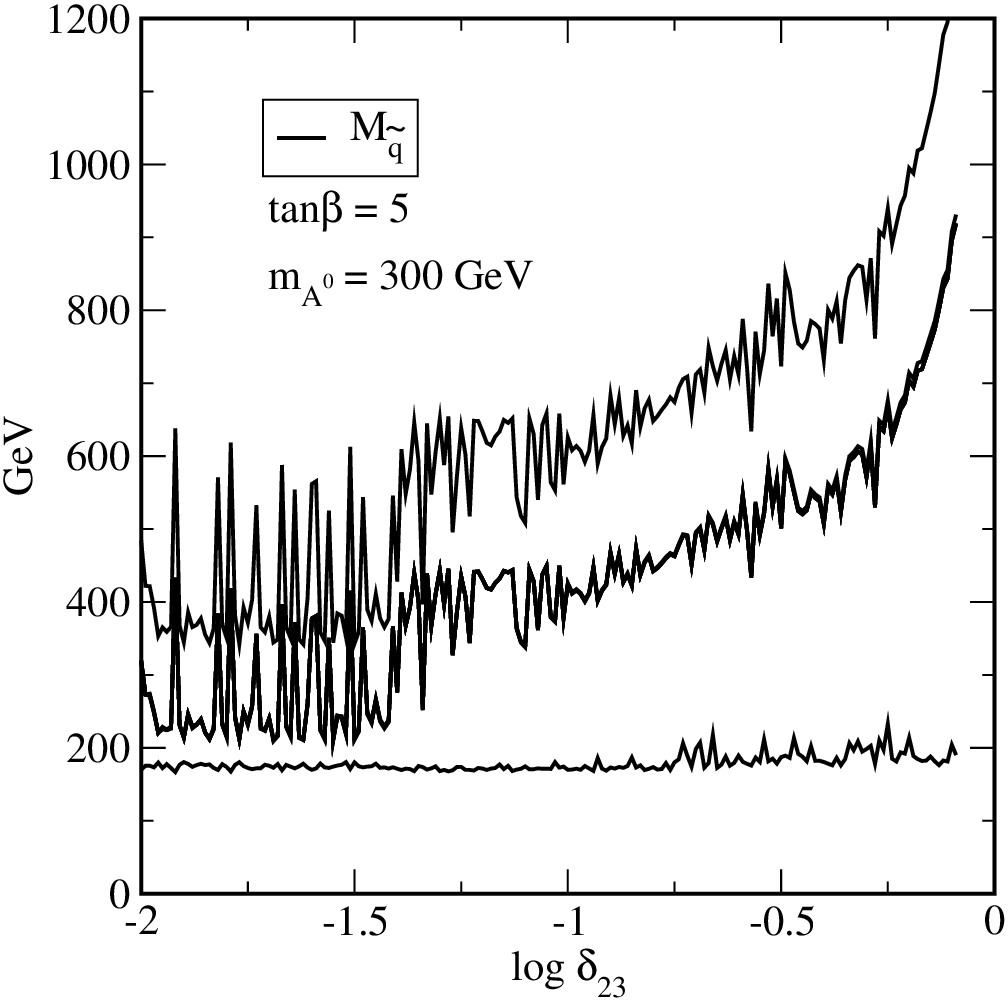} &
        \includegraphics*[height=0.45\textwidth]{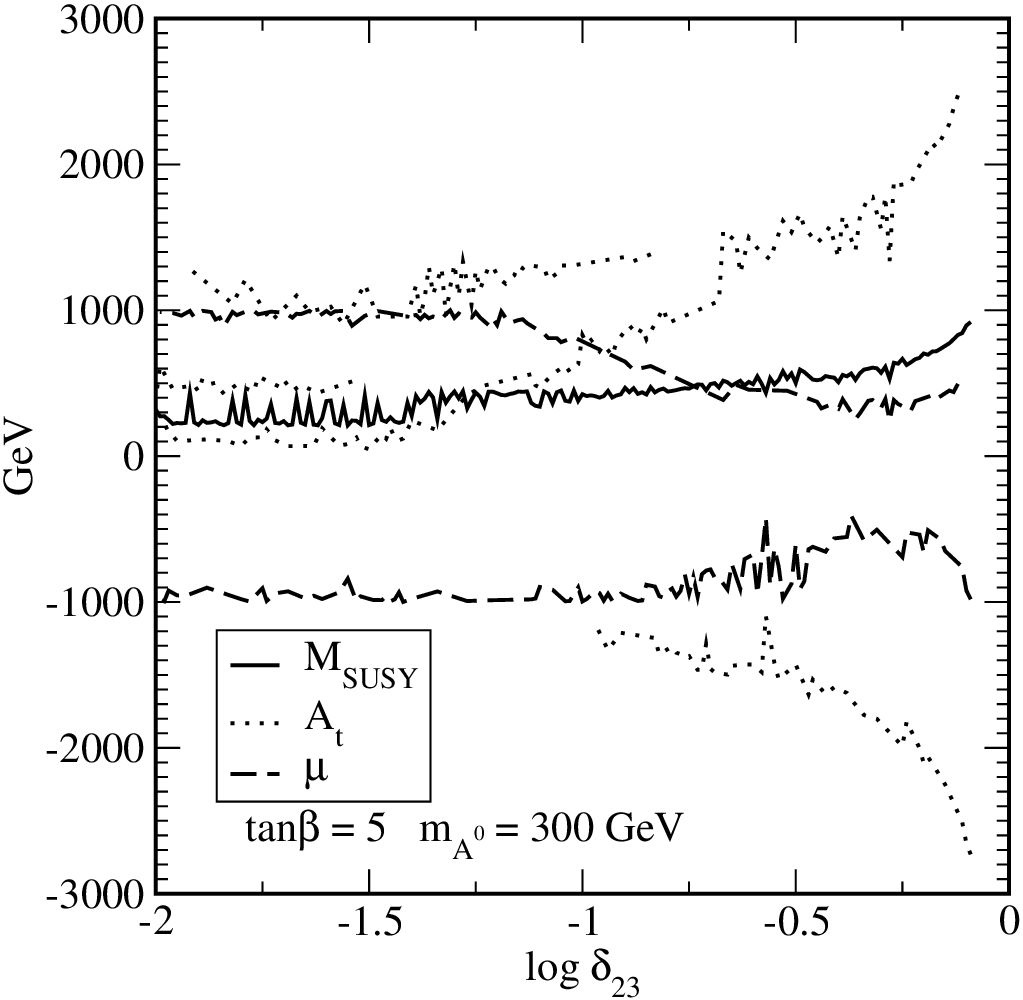} \\
        (a) & (b)
    \end{tabular}
    \caption{Value of \textbf{(a)} up-type squark physical  masses ($M_{\tilde{q}}$),
    four of them are degenerate;   \textbf{(b)} the
      parameters~(\ref{eq:scan-parameters}) from the maximization of
      the $H^0$ channel in Fig.~\ref{fig:htc-prod-bsg-d23}a.}
    \label{fig:htc-msq-param-d23}
\end{figure}

\begin{figure}[tp]
    \centering
    \includegraphics*[height=0.45\textwidth]{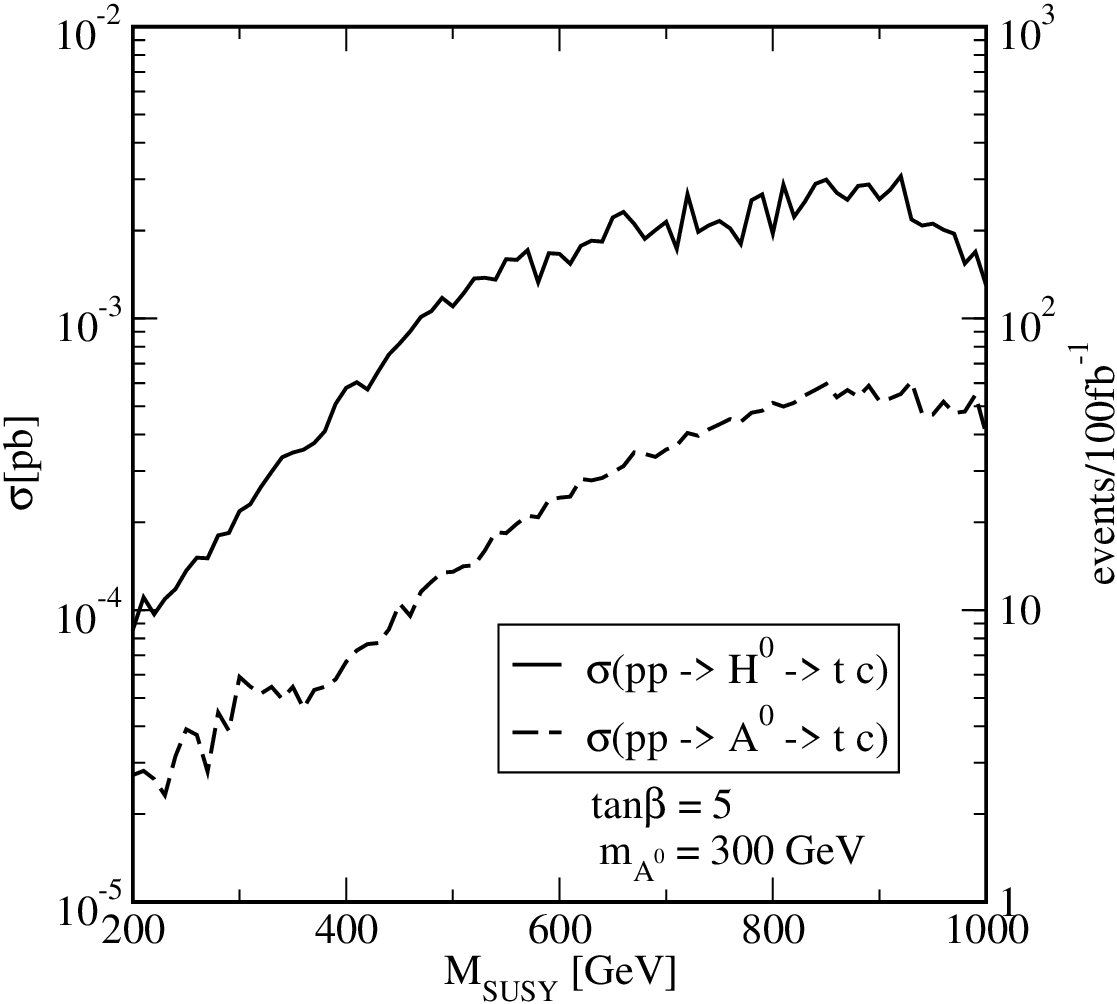}
    \caption{Maximum SUSY-QCD contributions to
      $\sigma(pp\rightarrow h\rightarrow t\,c)$, Eq.\,(\ref{eq:hqq-def}), as a
      function of \msusy.}
    \label{fig:htc-prod-msusy}
\end{figure}

We turn now our view to the role of the $\Bbsg$ restriction in
Figs.~\ref{fig:htc-prod-bsg-d23} and~\ref{fig:htc-msq-param-d23}.
Fig.~\ref{fig:htc-prod-bsg-d23} shows the maximum value of
$\sigmapphtc$ as a function of $\deltatt$ for a fixed value of
$\mA=300 \GeV$ together with the corresponding computed value of
$\Bbsg$, while Fig.~\ref{fig:htc-msq-param-d23} shows the values
of the parameters, Eq.~(\ref{eq:scan-parameters}), that realize
this maximum, together with the physical up-type squark masses.
In this case the $\Bbsg$ restriction is not as critical as in the
$bs$ channel, in part due to the fact that the SUSY-QCD
contribution to $\Bbsg$ is not enhanced at low $\tb$. For
$\deltatt\lesssim10^{-1.5}$ the value of $\Bbsg$ is well inside
the experimental limits (Fig.~\ref{fig:htc-prod-bsg-d23}b), there
is no restriction on the rest parameters
(Fig.~\ref{fig:htc-msq-param-d23}b), the squark masses remain
constant (Fig.~\ref{fig:htc-msq-param-d23}a), and the maximum
value of $\sigmapphtc$ grows here in the naively expected way
$(\deltatt)^2$ (Fig.~\ref{fig:htc-prod-bsg-d23}a).
Above this value ($\deltatt\gtrsim 10^{-1.5}$) the parameters
have to be adjusted to provide and acceptable range for
$\Bbsg$\,\footnote{The two lines appearing in this region for
$B_H(\bsg)$ mean that the maximum of $\sigmappHztc$ is attained
either by the maximum or the minimum allowed value of $\Bbsg$,
our Monte-Carlo sampling procedure picks either choice with equal
probability for each value of $\deltatt$.}. In this region the
SUSY mass $\msusy$ grows (Fig.~\ref{fig:htc-msq-param-d23}b), and
$|\mu|$ decreases, but not so fast as in the  $bs$ channel case
(Fig.~\ref{fig:hbs-msq-param-d23}b).
At the same time $A_t$ increases with increasing
$\deltatt> 10^{-1.5}$. Most of the physical squark masses grow,
but one of the up-type squarks (stop squark) can always have the
minimum allowed mass -- Eq.~(\ref{eq:scan-parameters}). In this
region the observables under study grow more slowly, since
their original $\deltatt^2$ behavior is partially
cancelled by the growing of $\msusy$ with $\deltatt$. In
Fig.~\ref{fig:htc-prod-msusy} we see, again, that $\sigmapphtc$
grows with the SUSY mass scale (although in a way less pronounced
than in the $bs$ channel), due to the relaxation of the $\bsg$
constraint for large $\msusy$. While $\sigmapphbs$ is augmented
nearly three orders of magnitude in the range
$\msusy=200-1000\GeV$ (Fig.~\ref{fig:hbs-prod-msusy}), the $tc$
channel undergoes only an increase of roughly a factor 10 in the
same parameter range.

How do the two Higgs boson channels $H^0$ and $A^0$ compare as
sources of $tc$ events? The $gg$-fusion mechanism is one of the
leading processes for Higgs boson production at relatively small
values of $\tb\gtrsim 1$ -- associated production with $b\bar{b}$
remaining still sizeable. Due to CP conservation, the squark
contributions to $gg\rightarrow A^0$ cancel out at one-loop and
only the quark contributions remain\,\cite{Spira:1997dg}. For
large squark masses the production cross-sections for $H^0$ and
$A^0$ are similar, the latter being slightly larger.  We see from
Fig.\,\ref{fig:htc-prod-ma-tb} that, despite the similarity in
production, the $H^0$ channel gives larger FCNC rates than the
$A^0$ one. The excess of FCNC events from the former can be
explained mainly from the constraints that we have imposed from
the very beginning on the squarks masses in relation to the Higgs
boson masses (see Eq.\,(\ref{eq:scan-parameters})). As we have
noted in section \ref{sec:numerical-analysis}, we can have
squarks of the same chirality-type in the $H^0\,\tilde{q}
\tilde{q}^*$ vertex, whereas they must necessarily be of opposite
chirality-type in the $A^0\tilde{q} \tilde{q}^*$ case. As a
result, for small $\mA$ the squark mass constraints expressed in
(\ref{eq:scan-parameters}) allow the FCNC cross-section
maximization process to pick points near the saturation of the
mass condition $2\,M_{\tilde{q_i}}>\mH+ 50\GeV$, but not of
$M_{\tilde{q}_i}+M_{\tilde{q}_j}>\mA + 50\GeV$ for $( i\neq j)$
because only one of the up-squarks can be light (see Fig.
\ref{fig:htc-msq-param-d23}a). This produces an enhancement of
the branching ratio of $H^0\to tc$ and for this reason this FCNC
channel dominates in the relatively small $\mA$ region. However,
as soon as $\mA$ is sufficiently heavy the second mass constraint
can also be satisfied and then the two curves in
Fig.~\ref{fig:htc-prod-ma-tb} tend to converge.

From the combined analysis of
Figs.~\ref{fig:htc-prod-ma-tb}-\ref{fig:htc-prod-msusy}, we
conclude that in the case of the $H^0$ channel we expect a maximum
of $\sim 300$ events/$100\fb^{-1}$ decays into top quarks at the
LHC. This maximum is achieved for a CP-odd Higgs boson mass of
$\mA\sim300 \GeV$, and a moderately low $\tb\sim5$. This rate can
grow one order of magnitude by a lower value of $\tb\sim2$, but
decreases significantly with $\mA$. The maximum is obtained at
the largest possible value of $\deltatt$, and a moderate SUSY
mass scale $\msusy\sim600-800\GeV$, but having one of the squarks
light. While the number of events is significantly lower than the
$bs$-channel ones, the $tc$-channels offers a better opportunity
for detection, due to the much lower background.

\section{Discussion and conclusions}
\label{sec:conclusions}

We have carried out a systematic study of the production rate of
FCNC processes at the LHC mediated by the decay of neutral Higgs
bosons of the MSSM: $\sigmapphqq\ (h=h^0,H^0,A^0)$ -- see
Eq.\,(\ref{eq:hqq-def}). Specifically, we have concentrated on
the FCNC production of the heavy quark pairs $qq'=bs$ and $tc$,
because they are the only ones that have a chance of being
detected. We have focused on the FCNC supersymmetric effects
stemming from the strongly interacting sector of the MSSM, namely
from the gluino-mediated flavor-changing interactions. We have
performed a maximization of the event rates in the parameter
space under a set of conditions that can be considered
``irreducible'', see Eq.\,(\ref{eq:scan-parameters}), i.e. we
cannot further shorten this minimal set (e.g. by making additional
assumptions on the relations among the parameters) without
potentially jeopardizing the conclusions of this study. Even
within this restricted parameter subspace the computer analysis
has been rather demanding. The numerical scan has been performed
using Monte Carlo techniques which we have partially
cross-checked with more conventional methods. The maximization of
the cross-sections (\ref{eq:hqq-def}) has been performed by
simultaneously computing the corresponding MSSM quantum effects
on the (relatively well-measured) low-energy FCNC decay $\bsg$
and requiring that the experimental limits on this observable are
preserved.

To summarize our results: the total number of FCNC heavy flavor
events originating from supersymmetric Higgs boson interactions at
the LHC can be large (of order $10^6$), but this does not mean
that they can be easily disentangled from the underlying
background of QCD jets where they are immersed. For example, it
is well known that the simple two-body decay $h\rightarrow
b\,\bar{b}$ is impossible to isolate due to the huge irreducible
QCD background from $b\,\bar{b}$ dijets -- a result that holds
for both the SM and the MSSM\,\cite{\LHC}. This led a long time
ago to complement the search with many other channels,
particularly $h\rightarrow\gamma\,\gamma$ which has been
identified as an excellent signature in the appropriate range.
Similarly, the FCNC Higgs boson decay channels may help to
complement the general Higgs boson search strategies, mainly
because the FCNC processes should be essentially free of QCD
background. Notwithstanding other difficulties can appear, such as
misidentification of jets. For instance, for the $bs$ final
states misidentification of $b$-quarks as $c$-quarks in
$cs$-production from charged currents may obscure the possibility
that the $bs$-events can be really attributed to Higgs boson FCNC
decays. This also applies to the $tc$ final states, where
misidentification of $b$-quarks as $c$-quarks in e.g. $tb$
production might be a source of background to the $tc$ events,
although in this case the clear-cut top quark signature should be
much more helpful (specially after an appropriate study of the
distribution of the signal versus the background).  However, to
rate the actual impact of these disturbing effects one would need
an additional study which is beyond the scope of the present
paper.

An interesting (and counter-intuitive) result of our work is that
for all the Higgs boson channels $h=h^0,H^0,A^0$, the FCNC
cross-section $\sigmapphqq$ increases with growing SUSY mass
scale $\msusy$. Due to this effective ``non-decoupling'' behavior
(which is more pronounced for the $bs$ channel) the FCNC rates
are maximal when the overall squark and gluino mass scale is of
order of $\msusy\lesssim 1\TeV$ -- with the only proviso that for
the $tc$-channel a single squark should have a low mass ($\gtrsim
150\GeV$).  Moreover, we find that the two types of FCNC final
states ($bs$ and $tc$) prefer different ranges of $\tb$. The
$bs$-channel is most efficient at high $\tb>30$, whereas the
$tc$-channel works better in the regime of low $\tb<10$ (see
below). As for the mixing parameter $\deltatt$ (which is the
fundamental supersymmetric FCNC parameter of our analysis, see
its definition in (\ref{delta23})) we remark that the maximum
number of events is not always attained for the largest possible
values of it (due to the influence of the $\bsg$ constraint): in
the $bs$-channel the maximum is achieved for moderate values in
the range $0.05\lesssim\deltatt\lesssim 0.2$, whereas the
$tc$-channel prefers the maximum allowed values in our analysis
($\deltatt\lesssim 0.8$). We also remark that the naive
expectation $\sigmapphqq\sim\deltatt^2$ does not always apply.

The number of FCNC events originating from the two channels $bs$
and $tc$ is not alike, and it also depends on the particular
Higgs boson. For the $bs$ final states we have found that the
optimized value of $\sigmapphbs$ produced by our analysis is
$\sim 12\pb$. This amounts to $\sim12 \times 10^5$ events per
$\int {\cal L}dt= 100\fb^{-1}$ of data at the LHC. The most
favorable Higgs boson channel is the one corresponding to the
lightest MSSM Higgs boson, $h^0$. For this boson there are
non-trivial correlations between the two factors in
Eq.~(\ref{eq:hqq-def}), namely between the Higgs boson production
cross-section and the FCNC branching ratio. These correlations
permit an increase of the total number of FCNC events up to two
orders of magnitude in certain cases as compared to the number of
events produced by the heavy Higgs bosons $H^0$ and $A^0$, which
are essentially free of these correlations. The latter stem from
relevant quantum effects on the parameters of the Higgs boson
sector at one-loop precisely in the regions of our interest.

On the other hand, the maximum value of $\sigmapphtc$ is more
moderate, to wit: $3\times 10^{-3}\pb$, or $\sim 300$
events/$100\fb^{-1}$. For the total integrated luminosity during
the operative lifetime of the LHC, which amounts to some
$(300-400)\fb^{-1}$, we estimate that a few thousand $tc$ events
could be collected in the most optimistic conditions. This number
is of course sensitive to many MSSM parameters, but most
particularly to two: $\mA$ and $\tb$. The mass of the CP-odd
Higgs boson should not be heavier than $\mA\sim (400-500)\GeV$ if
one does not want to decrease the number of events below a few
hundred per $100\fb^{-1}$. On the other hand the number of events
is very much dependent on the particular range of $\tb$. As we
have said above, the lowest possible values are preferred for the
$tc$ channel, but the sensitivity in this range is so high that
the order of magnitude of $\sigmapphtc$ may change for different
(close) choices of $\tb$. Throughout all the analysis of the $tc$
channel we have fixed $\tb$ at an intermediate value, but the
maximum number of events per $100\fb^{-1}$ would grow from $\sim
300$ (for our standard choice $\tb=5$) up to $\sim(500,900,2000)$
if we would have chosen $\tb=(4,3,2)$ respectively. Such lower
values of $\tb$ are usually avoided in some MSSM analyses in the
literature, but as a matter of fact there are no fully
water-tight experimental bounds on $\tb$ excluding this lower
range, apart from the more incontrovertible strict lowest limit
$\tb>1$. We recall that the lower limit on $\tb$ is obtained
  indirectly from the LEP exclusion data on the light Higgs boson
  search, and is therefore very sensitive to the
  inputs used in the computation, specially the top quark
  mass~\cite{Heinemeyer:1999zf}.
Unlike the difficulties in the $bs$-channel, and in spite of the
substantially smaller number of events, we deem more feasible to
extract the $tc$ signal at the LHC, due to the presence of the
quark top, which carries a highly distinguishable signature.

At this point a comparison with our previous
work\,\cite{Bejar:2003em} is in order. In that work we have
studied in quite some detail the maximum FCNC production rates of
Higgs bosons decaying into $tc$ final states within the general
two-Higgs-doublet model. It was found that the maximal branching
ratio in the 2HDM takes place in the type II 2HDM (or 2HDM II),
and reads $B^{II}(h\rightarrow tc)\sim10^{-5}$, whereas in the
2HDM I it is comparatively negligible. After a detailed
computation of the event rates (including also the particular
restrictions of the $\bsg$ process, which are different in the
2HDM case as compared to the MSSM) the conclusion was that
several hundred $tc$ events could be collected at the LHC under
optimal conditions. We clearly identified which are the most
relevant Higgs boson modes for this purpose and the domains of
the 2HDM II parameter space where these events could originate
from. To make it short, our conclusion there\,\cite{Bejar:2003em}
was that the $h^0$ is the most gifted decay and the ideal
situation occurs when $\tan\beta$ and $\tan\alpha$ are both
large, and also when the CP-odd state $A^0$ is much heavier than
the CP-even ones ($h^0,H^0$). Furthermore we found that in the
general 2HDM the $A^0$ state never gives any appreciable FCNC
rate into $tc$. It is easy to see that the mode $A^0\rightarrow
bs$ is not favored either (unless $\tb$ is very small and
$\mHp$ unusually light, both situations rather
unappealing).

How it compares with the MSSM case under study?  To start with, we
note that here the $A^0$ channel gives essentially the same $bs$
rate as the $H^0$ one, and that both modes can be quite relevant.
At the same time the $A^0$ rate into $tc$ is, though not dominant,
not negligible at all. Some few hundred events of this nature per
$100\fb^{-1}$ are possible. If this is not enough, in the MSSM
the most relevant $\tb$ region for the $tc$ final states is not
the highest one (as in the 2HDM II case) but just the opposite:
the lowest allowed one. This is an important difference, and one
that should help to discriminate between the 2HDM and MSSM models
in case that some $tc$ events would be unambiguously tagged at the
LHC. After all there are many high precision observables that are
highly sensitive to the preferred range of $\tb$, so that the
favorite value of $\tb$ could already be fixed from other
experiments by the time that some FCNC events could be detected.
Apart from the different correlation of the parameters in the
non-supersymmetric and supersymmetric model, in the latter case
the maximal event rates for the $tc$ mode are typically one order
of magnitude higher than the maximal event rates in the former.

The corresponding study for the 2HDM branching ratios into $bs$
final states was performed in~\cite{Arhrib:2004xu}, although in
this work the cross-section and number of events were not
computed. However, from the maximum size of the expected
branching ratios compatible with $\bsg$ (viz.
$B^{II}(h\rightarrow bs)\sim10^{-6}-10^{-5}$ in the 2HDM II) it
is already pretty obvious that the number of events can never be
competitive with the supersymmetric case where
$B^{MSSM}(h\rightarrow
bs)\sim10^{-4}-10^{-3}$\,\cite{Bejar:2004rz}. As for the 2HDM I
(which is insensitive to the $\bsg$ bounds) the branching ratios
can be at most of order $B^{I}(h\rightarrow bs)\sim
10^{-5}-10^{-3}$ and only so for very small values of
$\tb=0.1-0.5$ which are actually excluded in the MSSM.

Summing up and closing: the maximum number of FCNC events in the
MSSM case is larger than the highest expected rates both in the
2HDM I and II , the two kind of signatures of physics beyond the
SM being perfectly distinguishable because the relevant regions
of the parameter space are completely different. If a sample of
FCNC events of this kind could be collected, we should be able to
ascertain which is its ultimate origin.  At the end of the day if
one single thing should be emphasized is that the FCNC event rate
into $bs$ or $tc$ is so extremely tiny in the SM that if only a
dozen events of this kind could be captured under suitable
experimental conditions it would be an undeniable signature of
new physics. From what we have seen, the odds should be
heavily in favor of attributing it to a supersymmetric origin.

\section*{Acknowledgements}
The work of SB has been supported by CICYT under project
FPA2002-00648, by the EU network on Supersymmetry and the Early
Universe (HPRN-CT-2000-00152), and by DURSI Generalitat de
Catalunya under project 2001SGR-00188; JG  by a \textit{Ramon y
Cajal} contract from MEC (Spain); JG and JS in part by MEC and
FEDER under project 2004-04582-C02-01, and JS also by DURSI
Generalitat de Catalunya under project 2001SGR-00065.

\providecommand{\href}[2]{#2}
\bibliography{hbssqcd}

\end{document}